\documentclass[lettersize,journal]{IEEEtran}
\usepackage{amsmath,amsfonts,amssymb}
\usepackage{algorithmic}
\usepackage{algorithm}
\usepackage{tabularx}
\usepackage{array}
\usepackage{breqn}
\usepackage[caption=false,font=normalsize,labelfont=sf,textfont=sf]{subfig}
\usepackage{textcomp}
\usepackage{stfloats}
\usepackage{url}
\usepackage{verbatim}
\usepackage{graphicx}
\usepackage{cite}
\usepackage{bm}        
\usepackage{mathtools} 
\usepackage{amsthm}
\usepackage{booktabs}
\usepackage{arydshln}
\usepackage{pifont}
\usepackage{balance}
\usepackage{multirow}
\newcommand{\cmark}{\ding{51}}

\newtheorem{lem}{Lemma}
\newtheorem{pro}{Proposition}

\begin{document}

\title{Robust Transceiver Design for SIM-Assisted Integrated Sensing and Covert Communications}

\author{
Ahmed M. Benaya, \IEEEmembership{Member, IEEE}, Ali A. Nasir, \IEEEmembership{Senior Member, IEEE}, Khaled M. Rabie, \IEEEmembership{Senior Member, IEEE}, A. Abdelaziz Salem, and Daniel B. da Costa, \IEEEmembership{Senior Member, IEEE}
\thanks{This work is supported by the Deanship of Research at King Fahd University of Petroleum and Minerals (KFUPM) for funding under the Interdisciplinary Research Center for Communication Systems and Sensing (IRC-CSS).}
\thanks{Ahmed M. Benaya is with the Interdisciplinary Research Center for Communication Systems and Sensing (IRC-CSS), King Fahd University of Petroleum and Minerals (KFUPM), Dhahran 31261, Saudi Arabia. (e-mail: ahmeddiab@el-eng.menofia.edu.eg).}
\thanks{Ali A. Nasir and Daniel B. da Costa are with the Department of Electrical Engineering, King Fahd University of Petroleum and Minerals (KFUPM),  Dhahran 31261, Saudi Arabia, while Khaled M. Rabie is with the Department of Computer Engineering, King Fahd University of Petroleum and Minerals (KFUPM),  Dhahran 31261, Saudi Arabia. They all are also affiliated with the Center for Communication Systems and Sensing at KFUPM. (e-mail: anasir@kfupm.edu.sa, k.rabie@ieee.org, danielbcosta@ieee.org).}

\thanks{A. Abdelaziz Salem is with the College of Engineering and Technology, Fujairah University, UAE. He is also affiliated with the Department of Electronics and Electrical Communications Engineering, Faculty of Electronic Engineering, Menoufia University, Egypt. (e-mail: ahmed.abdalaziz40@el-eng.menofia.edu.eg).}}

\markboth{}{Ahmed. M. Benaya \MakeLowercase{\textit{et al.}}: Robust Transceiver Design for SIM-Assisted Integrated Sensing and Covert Communications}


\maketitle

\begin{abstract}
Stacked intelligent metasurfaces (SIMs) have emerged as a promising wave-domain processing technology for sixth-generation integrated sensing and communication networks, reducing reliance on complex radio-frequency hardware. Nevertheless, imperfect channel state information (CSI) can compromise sensing performance and communication reliability. Moreover, ensuring covert transmission introduces a further challenge, as the system must sense potential targets while concealing communication activity from unauthorized wardens. These challenges motivate the development of robust SIM-assisted integrated sensing and covert communication (ISACC) systems. In this paper, we investigate the robust transceiver design of a multi-user, multi-target ISACC system under imperfect CSI. We jointly optimize the transmit and receive beamformers, sensing covariance matrix, and SIM phase shifts to maximize the worst-case minimum sensing signal-to-interference-plus-noise ratio. The formulation enforces constraints on bounded channel uncertainties, user quality-of-service requirements, transmit-power budget, SIM phase shifts, and target detection probabilities. Since the formulated problem is highly non-convex, we develop an alternating optimization framework based on the S-procedure, successive convex approximation, and semidefinite relaxation. Furthermore, to reduce computational complexity, we derive exact lower-dimensional representations of the robust sensing linear matrix inequalities (LMIs) involved in transmit/receive-beamformer designs. We further develop a low-complexity phase-update method that avoids the large sensing LMIs. Numerical results demonstrate that the proposed design outperforms random phase shift and fully digital-beamforming benchmarks, while the low-complexity scheme achieves comparable performance to the LMI-based scheme at substantially lower computational cost.
\end{abstract}


\section{Introduction}
\IEEEPARstart{S}{ixth}-generation (6G) wireless networks are expected to support ultra-high data rate communication and advanced sensing services within a shared infrastructure, which requires flexible control of wireless signals and efficient use of hardware resources~\cite{Wang2023On}. Although large antenna arrays provide greater spatial degrees of freedom (DoF), fully digital implementations incur substantial hardware costs and power consumption due to the large number of radio frequency (RF) chains~\cite{li2017energy}. Stacked intelligent metasurfaces (SIMs) have emerged as a promising technology to address these challenges. A SIM consists of multiple programmable metasurface layers, each comprising numerous meta-atoms that manipulate electromagnetic waves as they propagate through the structure~\cite{An2023stacked, an2025stacked,liu2025multi}. By adjusting the phase shifts of individual meta-atoms, SIMs perform beamforming directly in the wave domain, offering fine control over signal propagation while significantly reducing reliance on complex RF hardware.

Such wave-domain processing capabilities can be particularly beneficial for integrated sensing and communication (ISAC). ISAC has emerged as a core paradigm for 6G networks to enable radar sensing and wireless communications over shared spectrum, hardware platforms, and transmitted waveforms~\cite{luo2026ISAC,wen2025asurvey}. Through wave-domain beamforming, a SIM can flexibly shape transmitted signals to simultaneously serve communication users and illuminate sensing targets\cite{niu2024stacked}. However, coordinating these dual functions requires balancing distinct operational requirements, where communication relies on high signal quality at user terminals, while sensing depends on the strength of target echoes relative to interference and noise. Moreover, since in multi-target scenarios, unwanted target echoes can severely degrade detection accuracy, joint transmit and receive beamforming provides an effective mechanism to manage this interference. Hence, sensing signal-to-interference-plus-noise ratio (SINR) serves as a suitable design metric that directly reflects the sensing performance~\cite{benaya2025aerial}.

In addition to these performance requirements, security is an essential consideration in ISAC systems. Since directing energy toward targets for sensing can increase their exposure to transmitted signals, such targets may also act as unauthorized observers that attempt to detect or intercept communication signals~\cite{alselwi2026sensing}. In contrast to conventional physical-layer security schemes that focus on degrading the received signal strength at malicious targets~\cite{salem2025robust}, covert communication aims to conceal the presence of communication activity through limiting the detection probability at malicious sensing targets~\cite{hu2025reconfigurable}. Incorporating covertness into SIM-assisted ISAC therefore requires careful control of the signals reaching both legitimate users and wardens. This task becomes more challenging under imperfect channel state information (CSI), where the channel estimation errors can weaken desired signals, increase interference, and cause violations of communication or covertness requirements. These challenges motivate a robust SIM-assisted ISAC design that jointly accounts for sensing, communication, and covertness under channel uncertainty.
\subsection{Related Work}
Stacked metasurfaces were first proposed in ~\cite{An2023stacked} to exploit multi-layer wave-domain processing and address the physical limitations of conventional single-layer architectures in order to significantly reduce RF chain requirements. Subsequent studies expanded the framework to account for discrete phase shifts and joint power allocation to maximize multi-user sum rates~\cite{an2025stacked,liu2025multi}. SIM-assisted communication has also been investigated in near-field and wideband scenarios. In particular, near-field beamfocusing designs exploited the additional distance-dependent spatial DoF to suppress inter-user interference~\cite{papazafeiropoulos2024near,benaya2026stacked}, while layer-by-layer SIM optimization was combined with digital precoding and power allocation to support multi-user transmission under frequency-selective channels and practical phase-tuning errors in wideband systems~\cite{li2025stacked}. In addition, to improve solution quality and reduce optimization cost, recent works have considered deep reinforcement learning (DRL)-based phase and power optimization, as well as refined alternating optimization (AO) and projected-gradient methods~\cite{bahingayi2025refined,liu2025multi}. Nevertheless, these works have mainly focused on rate-oriented metrics and relied on perfect CSI, leaving the critical vulnerability of channel estimation uncertainty largely unaddressed.

Recent studies have extended SIM technology to ISAC systems. A SIM-assisted downlink system with multiple communication users and a radar target has been investigated in~\cite{niu2024stacked}. They jointly optimized the SIM phase shifts and transmit power allocation to maximize the spectral efficiency under a sensing-power constraint. In~\cite{li2024transmit}, the authors considered a transmit beamforming design that balances the communication sum rate and the normalized sensing beampattern. They developed a dual-normalized differential gradient method to solve the resulting non-convex problem. A discrete phase shift-based SIM design has been studied along with transmit power optimization in~\cite{zhang2025joint}. Their AO approach uses variable separation and projection-based updates to manage multi-user interference and improve the sensing beam gain. In~\cite{wang2024multi}, Wang \textit{et al.} have considered multi-user SIM-assisted ISAC with an extended target, where they jointly optimized the transmit beamformer and the end-to-end SIM response to minimize the Cramér-Rao bound under user-SINR and transmit-power constraints. Their method combines AO and semidefinite relaxation (SDR) and is also supported by experimental results. Although these studies demonstrate the potential of SIMs for ISAC, they have not explicitly considered CSI uncertainty errors or the joint transmit/receive beamforming design.

By providing additional wave-domain DoF without requiring extra RF chains, SIMs can improve secure transmission by directing information signals toward intended users and controlling leakage toward unintended receivers. Building on this capability, the works in~\cite{niu2024enhancing,niu2024efficient} investigate secure single-input-single-output (SISO) transmission and jointly design the SIM phase shifts, transmit signals, and artificial noise to enhance secrecy performance. These secrecy-based methods protect the message content by limiting an eavesdropper’s ability to decode it. Covert communication provides a different form of protection by hiding the existence of the transmission itself. This capability is especially important in ISAC, where sensing signals illuminate potential targets that may also monitor the communication activity. Along this direction, \cite{xie2026covert} studies covert ISAC with SIMs and uses DRL to jointly optimize the transmit power and SIM phase shifts under sensing and covert-communication constraints. In a complementary study, \cite{zhang2024robust} develops a robust covert ISAC system using a fully digital BS without any configurable surface assistance. Although the former does not consider CSI uncertainty, the latter does not exploit SIM-enabled wave-domain processing, leaving robust SIM-assisted covert ISAC largely unexplored. 

\subsection{Motivation and Main Contributions}
\begin{table*}[t]
\centering
\caption{Comparison of Our Work with Existing SIM-Assisted and ISAC Wireless Systems}
\label{tab:comparison}
\setlength{\tabcolsep}{4.5pt}
\renewcommand{\arraystretch}{1.1}
\begin{tabular}{l cccccccccccc}
\toprule
\textbf{Contribution} & \cite{niu2024stacked} & \cite{papazafeiropoulos2024near} &\cite{benaya2026stacked}& \cite{li2025stacked}& \cite{bahingayi2025refined} & \cite{li2024transmit} & \cite{zhang2025joint}& \cite{wang2024multi} & \cite{xie2026covert} & \cite{zhang2024robust} & \cite{ul2026robust} & \textbf{Our paper} \\
\midrule
Wave-domain processing  & \cmark & \cmark &\cmark&\cmark & \cmark & \cmark &\cmark & \cmark & \cmark &  & \cmark & \cmark\\
\hdashline
Multi-user communication  & \cmark & \cmark &\cmark &\cmark& \cmark & \cmark &\cmark& \cmark & \cmark & \cmark & \cmark & \cmark\\
\hdashline
ISAC functionality  & \cmark & &&& & \cmark &\cmark & \cmark & \cmark & \cmark &  & \cmark \\
\hdashline
Multi-target sensing  &  & && & & \cmark & &  & \cmark & \cmark &  & \cmark \\
\hdashline
Covert communication  & & &\cmark&& & & & & \cmark & \cmark &  & \cmark\\
\hdashline
Imperfect CSI / Robust design & & && & & & & & & \cmark & \cmark & \cmark \\
\hdashline
Joint Tx and Rx beamforming & & &&&& & & & & \cmark & & \cmark \\
\hdashline
Low-complexity / Dimension-reduction & & & & &  & & & & & & & \cmark \\
\bottomrule
\end{tabular}
\end{table*}
The reviewed literature reveals several important limitations. Existing SIM-assisted ISAC studies are mainly developed for nominal CSI and use sensing power, beampattern gain, or the Cramér-Rao bound as sensing metrics. They do not jointly optimize the transmit and receive beamformers to suppress interference among multiple sensing targets. Moreover, the security-oriented SIM studies have considered secrecy in simple SISO links or covert ISAC using DRL. More recently, robust multi-user downlink beamforming for SIMs with imperfect CSI and finite-resolution phase shifts was investigated in~\cite{ul2026robust}, whereas robust covert ISAC using fully digital beamforming was considered in~\cite{zhang2024robust}. The former does not include sensing or security constraints, while the latter does not exploit metasurface-assisted wave-domain processing. Hence, the joint design of SIM-assisted sensing, communication, and covertness under channel uncertainty remains largely unexplored. Although integrating SIM into covert ISAC significantly improves secure communication and sensing performance, it poses severe mathematical challenges in guaranteeing quality-of-service (QoS) and covertness due to the tight coupling among multi-layer phase shifts, digital transmit/receive beamformers, and CSI estimation errors. Compared to the literature, the key contributions of our work, summarized in Table~\ref{tab:comparison}, are detailed as follows:

\begin{itemize}
\item We propose a robust transceiver design framework for a multi-user, multi-target SIM-assisted ISACC system under imperfect CSI. By jointly optimizing the transmit/receive beamformers, sensing covariance matrix, and SIM phase shifts, the proposed design maximizes the worst-case minimum sensing SINR while satisfying communication QoS, transmit power budgets, SIM phase constraints, and warden detection-probability bounds.

\item To tackle the severe non-convexity caused by CSI uncertainties and the tight coupling among transmit/receive beamformers and multi-layer SIM phase shifts, we develop an AO framework, where the original problem is decoupled into three sub-problems; receive beamforming, transmit beamforming, and SIM phase shift optimization. The S-procedure is leveraged to convert the bounded channel uncertainties into tractable linear matrix inequalities (LMI) deterministic constraints.

\item We derive exact lower-dimensional representations of the robust sensing LMIs arising in the transmit- and receive-beamformer sub-problems, thereby reducing their computational dimensions without introducing additional approximation. Moreover, we propose a low-complexity SIM phase-update method based on norm bounds that substantially reduces the computational cost associated with the LMI-based phase-update formulation.

\item Numerical results show that the proposed design outperforms random phase shift and conventional digital-beamforming benchmarks. They also demonstrate that the low-complexity phase-update method achieves performance close to the LMI-based design with considerably lower computational complexity.
\end{itemize}

The remainder of the paper is organized as follows: the SIM-assisted multi-user multi-target ISACC system model, communication and sensing models, covertness metric, and channel model are presented in Section~\ref{sec_sys}. Section~\ref{sec_formulation} presents the robust max-min sensing SINR problem formulation. The proposed AO problem solution is shown in~\ref{sec_opti}. Simulation results and discussions are presented in Section~\ref{results}. Finally, Section~\ref{sec_conc} concludes the paper.

\section{System Model}
\label{sec_sys}

\begin{figure}[htpb]
\centerline{\includegraphics[width=0.5\textwidth]{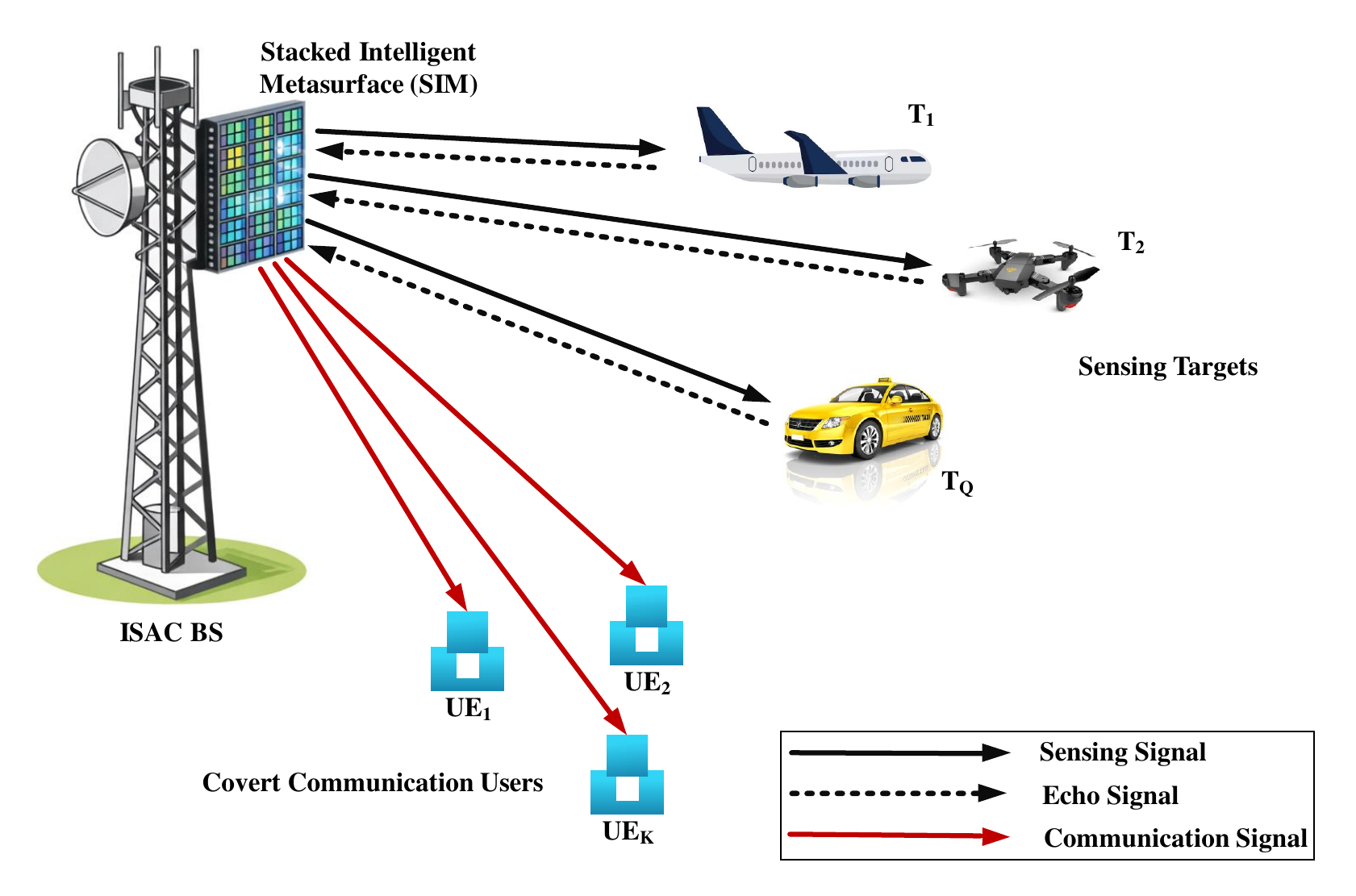}}
\caption{SIM-assisted multi-user multi-target integrated sensing and covert communication system.}
\label{fig_1}
\end{figure}
As shown in Fig.~\ref{fig_1}, we consider a SIM-enabled ISACC system, 
where a BS equipped with a uniform linear array (ULA) of $N_t$ transmit antennas and $N_r$ receive antennas serving $K$ single-antenna covert users while simultaneously sensing $Q$ potential eavesdropping targets. A SIM composed of $L$ layers is deployed in front of the BS to perform wave-domain beam manipulation for both communication and sensing functionalities, where each layer contains $M$ passive meta-atoms. Let \( \mathcal{K} = \{1,2,\ldots,K\} \), 
\( \mathcal{Q} = \{1,2,\ldots,Q\} \), \( \mathcal{N}_t = \{1,2,\ldots,N_t\} \),  \( \mathcal{N}_r = \{1,2,\ldots,N_r\} \),
\( \mathcal{L} = \{1,2,\ldots,L\} \), and 
\( \mathcal{M} = \{1,2,\ldots,M\} \) denote the index sets of covert UEs, eavesdropping targets, transmit antennas, receive antennas, SIM layers, and meta-atoms per layer, respectively. 
\subsection{SIM Model}
Each layer of the stacked intelligent metasurface (SIM) is configured as a rectangular uniform planar array (UPA) with 
\( M = M_x \times M_z \) elements, 
where \( M_x \) and \( M_z \) denote the numbers of elements along the \(x\)- and \(z\)-axes, respectively. 
Based on the Rayleigh--Sommerfeld diffraction formulation~\cite{lin2018all,liu2022programmable}, 
the propagation coefficient between the \( \tilde{m} \)-th element of the \((l-1)\)-th layer and the \(m\)-th element of the \(l\)-th layer, 
for all \( m, \tilde{m} \in \mathcal{M} \) and \( l \in \mathcal{L} \), is given by
\begin{equation}
\omega_{m,\tilde{m}}^l =
\frac{A_e \cos\!\left(\chi^l_{m,\tilde{m}}\right)}{r^l_{m,\tilde{m}}}
\left(
\frac{1}{2\pi r^l_{m,\tilde{m}}}
- j\frac{1}{\lambda}
\right)
e^{j2\pi r^l_{m,\tilde{m}}/\lambda},
\label{Eq:1}
\end{equation}
where \( A_e \) denotes the area of each element, 
\( \chi^l_{m,\tilde{m}} \) is the angle between the propagation direction and the surface normal of the \((l-1)\)-th layer, 
\( r^l_{m,\tilde{m}} \) represents the distance between the two elements, 
and \( \lambda \) is the carrier wavelength.

The complex reflection coefficient of the \(m\)-th element in the \(l\)-th layer is represented as 
\(e^{j\theta_m^l} \), 
for all \( m \in \mathcal{M} \) and \( l \in \mathcal{L} \), where \( \theta_m^l \in [0,2\pi) \) denote the reflection phase shift. 
Accordingly, the diagonal phase shift matrix of the \(l\)-th metasurface layer is 
\begin{equation}
\mathbf{\Theta}^l = 
\mathrm{diag}\!\left(
e^{j\theta_1^l}, 
e^{j\theta_2^l}, 
\dots, 
e^{j\theta_M^l}
\right)
\in \mathbb{C}^{M \times M}, 
\quad \forall l \in \mathcal{L}.
\label{Eq:2}
\end{equation}

Consequently, the overall forward SIM transfer matrix is
\begin{equation}
\mathbf{G}
=
\mathbf{\Theta}^L \mathbf{\Omega}^L
\cdots
\mathbf{\Theta}^2 \mathbf{\Omega}^2
\mathbf{\Theta}^1 \mathbf{\Omega}_{t}^1
\in \mathbb{C}^{M \times N_t},
\label{Eq:3}
\end{equation}
where \( \mathbf{\Omega}^l \in \mathbb{C}^{M \times M} \) for all \( l \in \mathcal{L} \setminus \{1\} \) 
denotes the inter-layer propagation matrix from the \((l-1)\)-th layer to the \(l\)-th layer. 
The matrix \( \mathbf{\Omega}_t^1 \in \mathbb{C}^{M \times N_t} \), which models the channel between the transmit antenna array and the first SIM layer, 
can be derived from~\eqref{Eq:1} by replacing 
\( r^l_{m,\tilde{m}} \) and \( \chi^l_{m,\tilde{m}} \) 
with \( r^1_{m,n_t} \) and \( \chi^1_{m,n_t} \), respectively, 
for all \( m \in \mathcal{M} \) and \( n_t \in \mathcal{N}_t \).

\subsection{Communication and Sensing Models}
We consider a narrowband flat-fading downlink communication model and a narrowband monostatic radar model over a transmission block consisting of $T$ symbol intervals, where $\mathcal{T}=\{1,2,\ldots,T\}$ denotes the index set of symbol times. Let $\mathbf{x}[t]\in\mathbb{C}^{N_t\times 1}$ denote the complex-baseband transmit vector at symbol time $t$.
The BS transmits $K$ communication streams and a dedicated sensing signal $\mathbf{x}_0[t]\in\mathbb{C}^{N_t\times 1}$, which is modeled as an independent and identically distributed (i.i.d.) zero-mean complex Gaussian random vector across time slots, with covariance matrix $\mathbf R_0 \succeq 0$, i.e., $x_0[t] \sim \mathcal{CN}(0, \mathbf R_0)$.

Let $\mathbf{s}[t] \triangleq\left [s_1\left[t\right], s_2\left[t\right], \dots, s_K\left[t\right]\right ] \in\mathbb{C}^{K\times 1}$ be the i.i.d. zero-mean complex Gaussian communication symbol vector with $\mathbb{E}\!\left\{\mathbf{s}[t]\mathbf{s}[t]^H\right\}=\mathbf{I}_K$
and $\mathbf{w}_k\in\mathbb{C}^{N_t\times 1}$ denote the digital communication beamforming vector toward the $k$-th covert user. Hence,
The transmit signal is modeled as
\begin{equation}
\mathbf{x}[t] = \sum_{k\in\mathcal{K} }\mathbf{w}_k s_k[t] + \mathbf{x}_0[t].
\label{eq:4}
\end{equation}

Assuming sufficiently large $T$, the transmit covariance matrix can be expressed as
\begin{equation}
\mathbf{R}_x \triangleq \mathbb{E}\!\left\{\mathbf{x}[t]\mathbf{x}[t]^H\right\}
= \sum_{k\in\mathcal{K} } \mathbf{w}_k \mathbf{w}_k^H + \mathbf{R}_0,
\label{eq:5}
\end{equation}
where $\mathbf{R}_0 = \frac{1}{T} \sum_{t \in \mathcal{T}} \mathbf{x}_0[t]\mathbf{x}_0[t]^H$. Accordingly, the total power budget constraint is expressed as
\begin{equation}
\mathrm{Tr}(\mathbf{R}_x)=\sum_{k\in\mathcal{K}} \|\mathbf{w}_k\|^2+\mathrm{Tr}(\mathbf{R}_0) \le P_{\max},
\label{eq:6}
\end{equation}
where $P_{\max}$ is the maximum allowed transmit power at the BS.

Let $\mathbf{h}_k\in\mathbb{C}^{M\times 1}$ denote the channel from the outermost SIM layer to covert user $k\in\mathcal{K}$.
The received signal at user $k$ can be written as
\begin{equation}
y_k[t] = \mathbf{h}_k^H \mathbf{G}\mathbf{x}[t] + n_k[t],
\label{eq:7}
\end{equation}
where $n_k[t]\sim\mathcal{CN}(0,\sigma_k^2)$ denotes the additive white Gaussian noise (AWGN) component at the $k$-th covert user. By substituting \eqref{eq:4} into \eqref{eq:7}, we obtain
\begin{dmath}
y_k[t]
=
\underbrace{\mathbf{h}_k^H\mathbf{G}\mathbf{w}_k s_k[t]}_{\text{desired}}
+
\underbrace{\sum_{i\in\mathcal{K},i\neq k}\mathbf{h}_k^H\mathbf{G}\mathbf{w}_i s_i[t]}_{\text{multi-user interference}}
+
\underbrace{\mathbf{h}_k^H\mathbf{G}\mathbf{x}_0[t]}_{\text{sensing leakage}}
+ n_k[t].
\label{eq:yk_expand}
\end{dmath}

Then, the downlink SINR at user $k$ is given by
\begin{equation}
\gamma_k
=
\frac{|\mathbf{h}_k^H\mathbf{G}\mathbf{w}_k|^2}
{\sum_{i\in\mathcal{K},i\neq k}|\mathbf{h}_k^H\mathbf{G}\mathbf{w}_i|^2
+\mathbf{h}_k^H\mathbf{G}\mathbf{R}_0\mathbf{G}^H\mathbf{h}_k
+\sigma_k^2 }.
\label{eq:sinr_comm}
\end{equation}

Let $\mathbf{a}_q\in\mathbb{C}^{M\times 1}$ denote the channel from the SIM's outermost layer to the target $q\in\mathcal{Q}$.
The effective forward channel from the BS transmit array to target $q$ is
\begin{equation}
\mathbf{f}_{1,q} \triangleq \mathbf{a}_q^H\mathbf{G}\in\mathbb{C}^{1\times N_t}.
\label{eq:g1q}
\end{equation}

To model the return path to the $N_r$-antenna BS receiver, we introduce a receive-side wave-domain transfer matrix
$\mathbf{B}\in\mathbb{C}^{M\times N_r}$ from the BS receive array to the SIM aperture in the reverse direction.
The effective return channel from target $q$ to the BS receive array is
\begin{equation}
\mathbf{f}_{2,q} \triangleq \mathbf{B}^H \mathbf{a}_q \in\mathbb{C}^{N_r\times 1}.
\label{eq:g2q}
\end{equation}

Consequently, the resulting round-trip MIMO channel matrix associated with the $q$-th target is expressed as
\begin{equation}
\mathbf{F}_q \triangleq \alpha_q\,\mathbf{f}_{2,q}\mathbf{f}_{1,q} = \alpha_q\,\mathbf{B}^H \mathbf{A}_q\mathbf{G}\in \mathbb{C}^{N_r\times N_t}, 
\label{eq:Gq}
\end{equation}
where $\alpha_q\in\mathbb{C}$ denotes the scattering coefficient of the $q$-th target and $\mathbf{A}_q=\mathbf{a}_q\mathbf{a}_q^H$ is the round-trip channel between the SIM and the $q$-th target. Accordingly, the received echo signal at the BS is
\begin{equation}
\mathbf{r}[t]
=
\sum_{q\in\mathcal{Q}} \mathbf{F}_q \mathbf{x}[t] + \mathbf{n}_B[t],
\label{eq:r}
\end{equation}
where $\mathbf{n}_B[t]\sim\mathcal{CN}(\mathbf{0},\sigma_B^2\mathbf{I}_{N_r})$ is the AWGN component at the BS.

Let $\mathbf{u}_q\in\mathbb{C}^{N_r\times 1}$ denote the receive beamformer for extracting target $q$.
The corresponding observation is
\begin{equation}
\hat{r}_q[t] = \mathbf{u}_q^H \mathbf{r}[t].
\label{eq:rhat}
\end{equation}

Treating echoes from other targets as interference, the sensing SINR for target $q$ is defined as
\begin{equation}
\gamma_q
=
\frac{\mathbf{u}_q^H\mathbf{F}_q \mathbf{R}_x \mathbf{F}_q^H \mathbf{u}_q}
{\sum_{j\in\mathcal{Q},j\neq q}\mathbf{u}_q^H\mathbf{F}_j \mathbf{R}_x \mathbf{F}_j^H \mathbf{u}_q
+\sigma_B^2\|\mathbf{u}_q\|^2 }.
\label{eq:sinr_sens}
\end{equation}

\subsection{Covertness Metric and Detection Model}

Following \cite{zhou2025star,liu2024ris}, we adopt the detection error probability (DEP) as the metric to quantify covertness.
Assuming equal prior probabilities, the DEP at target $q\in\mathcal{Q}$ is $\xi_q \triangleq P_{\mathrm{FA},q} + P_{\mathrm{MD},q}$, where $P_{\mathrm{FA},q}$ and $P_{\mathrm{MD},q}$ denote the false-alarm and missed-detection probabilities, respectively. The signal received at the $q$-th target in the $t$-th symbol interval is modeled as
\begin{equation}
y_{q}[t]=
\begin{cases}
\mathbf{f}_{1,q}\mathbf{x}_0[t]+n_{q}[t], & \mathcal{H}_0,\\[2mm]
\sum_{k\in\mathcal{K} } \mathbf{f}_{1,q} \mathbf{w}_k s_k[t] + \mathbf{f}_{1,q}\mathbf{x}_0[t]+n_{q}[t], & \mathcal{H}_1,
\end{cases}
\label{eq:warden_obs}
\end{equation}
where $\mathcal{H}_0$ represents the null hypothesis (no covert transmission) and $\mathcal{H}_1$ denotes the alternative hypothesis (BS transmits covert signals). Moreover, $n_{q} [t]\sim \mathcal{CN}(0,\sigma_{q}^2)$ denotes the AWGN component at the $q$-th target. Stacking the $T$ received signals over the transmission block yields the observation vector $\mathbf{y}_q \triangleq \big[y_{q}[1],y_{q}[2],\ldots,y_{q}[T]\big]^T$. Since the elements of $\mathbf{y}_q$ are i.i.d.\ circularly-symmetric complex Gaussian random variables under each hypothesis, the likelihood functions are given by
\begin{equation}
\begin{aligned}
\mathbb{P}_{1,q} &\triangleq p_{\mathcal{H}_1,q}\!\left(\mathbf{y}_q\right)
= \frac{1}{\pi^T \delta_{1,q}^T}
\exp\!\left(-\frac{\mathbf{y}_q^H\mathbf{y}_q}{\delta_{1,q}}\right),\\
\mathbb{P}_{0,q} &\triangleq p_{\mathcal{H}_0,q}\!\left(\mathbf{y}_q\right)
= \frac{1}{\pi^T \delta_{0,q}^T}
\exp\!\left(-\frac{\mathbf{y}_q^H\mathbf{y}_q}{\delta_{0,q}}\right),
\end{aligned}
\label{eq:likelihoods}
\end{equation}
where the variances $\delta_{0,q} = \mathbf{f}_{1,q}\mathbf{R}_0\mathbf{f}_{1,q}^H + \sigma_{q}^2$ and $\delta_{1,q} = \sum_{k\in\mathcal{K} } |\mathbf{f}_{1,q} \mathbf{w}_k|^2+\mathbf{f}_{1,q}\mathbf{R}_0\mathbf{f}_{1,q}^H + \sigma_{q}^2$. Let $\xi_q^\star$ denote the minimum achievable DEP at the $q$-th target with the optimal detector.
As shown in \cite{yan2019gaussian}, it satisfies
\begin{equation}
\xi_q^\star = 1-\mathcal{V}_q\!\left(\mathbb{P}_{0,q},\mathbb{P}_{1,q}\right),
\label{eq:dep_tvd}
\end{equation}
where $\mathcal{V}_q(\cdot,\cdot)$ is the total variation distance (TVD).
Applying Pinsker's inequality \cite{yan2019gaussian,zhang2024robust}, the TVD can be upper-bounded by the Kullback--Leibler (KL) divergence as
\begin{equation}
\mathcal{V}_q\!\left(\mathbb{P}_{0,q},\mathbb{P}_{1,q}\right)
\le
\sqrt{\frac{\mathcal{D}_q\!\left(\mathbb{P}_{1,q}\Vert\mathbb{P}_{0,q}\right)}{2}},
\label{eq:pinsker}
\end{equation}
where $\mathcal{D}_q(\mathbb{P}_{1,q}\Vert\mathbb{P}_{0,q})$ denotes the KL divergence.
For the i.i.d.\ complex Gaussian model in \eqref{eq:likelihoods}, the KL divergence admits the closed-form expression
\begin{equation}
\mathcal{D}_q\!\left(\mathbb{P}_{1,q}\Vert\mathbb{P}_{0,q}\right)
=
T\,\nu\!\left(\frac{\delta_{1,q}-\delta_{0,q}}{\delta_{0,q}}\right), ~~
\nu(x)\triangleq x-\ln(1+x).
\label{eq:kl_closed_form}
\end{equation}

Therefore, ensuring a prescribed covertness level can be achieved by imposing
\begin{equation}
\xi_q^\star \ge 1-\epsilon
\ \Longleftarrow\
\mathcal{D}_q\!\left(\mathbb{P}_{1,q}\Vert\mathbb{P}_{0,q}\right)\le 2\epsilon^2,
\quad \forall q\in\mathcal{Q},
\label{eq:covert_constraint_kl}
\end{equation}
where $\epsilon\in(0,1)$ is a small constant controlling the desired covertness requirement.

\subsection{Channel Model with Bounded CSI Uncertainty}
Rician fading is adopted to model the channels from the outermost SIM layer to the $k$-th covert user, $\mathbf{h}_k$, i.e.,
\begin{equation}
\mathbf{h}_k
=
\sqrt{l_k}\left(\sqrt{\frac{\kappa}{\kappa+1}}\,\bar{\mathbf{h}}_k
+
\sqrt{\frac{1}{\kappa+1}}\,\tilde{\mathbf{h}}_k\right),
\quad \forall k\in\mathcal{K},
\label{eq:rician_hk}
\end{equation}
where $\kappa\ge 0$ is the Rician factor and $l_k = l_0 d_k^{-\eta}$ is the large-scale fading coefficient with $d_k$ denotes the propagation distance, where $l_0 = \left(\frac{\lambda}{4\pi}\right)^2$ is the free space path loss at a reference distance $1$m and $\eta$ is the path loss exponent. In addition, $\bar{\mathbf{h}}_k$ denotes the deterministic line-of-sight (LoS) component, and
$\tilde{\mathbf{h}}_k$ is the non-LoS component with $\tilde{\mathbf{h}}_k \sim \mathcal{CN}\!\left(\mathbf{0},\mathbf{R}_{h_k}\right)$, where $\mathbf{R}_{h_k}\in\mathbb{C}^{M\times M}$ is the spatial correlation matrix that quantifies the correlation among the channels associated with different meta-atoms with $[\mathbf{R}_{h_k}]_{m,m'}=\textrm{sinc}\left( \frac{2\delta_{m,m'}}{\lambda}\right)$, where $\delta_{m,m'}$ is the corresponding meta-atom spacing.

As stated previously, the SIM lies in the $x$--$z$ plane and consists of $M = M_x \times M_z$ elements. Here, $M_x = 2\tilde{M}_x + 1$ denotes the number of elements along the $x$-axis, and $M_z = 2\tilde{M}_z + 1$ denotes the number of elements along the $z$-axis. The corresponding element spacings in the $x$ and $z$ directions are $d_x$ and $d_z$, respectively. When the user lies in the far field of the SIM aperture, the LoS component $\bar{\mathbf{h}}_k$ can be represented according to~\cite{liu2023near} with $\bar{\mathbf{h}}_k = \mathbf{a}_x(\vartheta_k,\varphi_k)\otimes \mathbf{a}_z(\vartheta_k,\varphi_k)\in\mathbb{C}^{M\times 1}$, where $(\vartheta_k,\varphi_k)$ denote the azimuth and elevation angles of the $k$-th covert user, respectively. Moreover, the array responses $\mathbf{a}_x(\vartheta_k,\varphi_k)$ and $\mathbf{a}_z(\vartheta_k,\varphi_k)$ are given by

\small
\begin{dmath}
    \mathbf{a}_x(\vartheta_k,\varphi_k)
=
\left[
\,
e^{-j \frac{2\pi}{\lambda} \tilde{M}_x d_x \cos\vartheta_k\sin\varphi_k},\,
\dots,\,
e^{j \frac{2\pi}{\lambda} \tilde{M}_x d_x \cos\vartheta_k\sin\varphi_k}
\right]^T,
\label{eq:ax}
\end{dmath}
and
\begin{dmath}
\mathbf{a}_z(\vartheta_k,\varphi_k) =
\left[
\,
e^{-j \frac{2\pi}{\lambda} \tilde{M}_z d_z \cos\varphi_k},\,
\dots,\,
e^{j \frac{2\pi}{\lambda} \tilde{M}_z d_z \cos\varphi_k}
\right]^T.
\label{eq:az}
\end{dmath}
\normalsize

Similarly, the SIM-to-target channel can be modeled as
\begin{equation}
\mathbf{a}_q
= \sqrt{l_q} \left(
\mathbf{a}_x(\vartheta_q,\varphi_q)\otimes \mathbf{a}_z(\vartheta_q,\varphi_q)\right)\in\mathbb{C}^{M\times 1},
\quad \forall q\in\mathcal{Q},
\label{eq:aq_array}
\end{equation}
where $\mathbf{a}_x(\vartheta_q,\varphi_q)$ and $\mathbf{a}_z(\vartheta_q,\varphi_q)$ can be expressed using~\eqref{eq:ax} and~\eqref{eq:az} with $(\vartheta_q,\varphi_q)$ denoting the departure angles from the SIM towards the $q$-th target. In addition, $l_q = l_0 d_q^{-\eta}$ is the large-scale fading coefficient, where $d_q$ denotes the propagation distance towards the $q$-th target.

To account for CSI imperfections, we adopt the bounded CSI model.
Specifically, for each user $k$ and target $q$, the actual channels are expressed as
\begin{align}
\mathbf{h}_k &= \hat{\mathbf{h}}_k + \Delta\mathbf{h}_k,
\qquad \,\,
\|\Delta\mathbf{h}_k\|_2 \le \varepsilon_{h_k},
\quad \forall k\in\mathcal{K},
\label{eq:hk_unc}
\\
\mathbf{a}_q &= \hat{\mathbf{a}}_q + \Delta\mathbf{a}_q,
\qquad\,\,~
\|\Delta\mathbf{a}_q\|_2 \le \varepsilon_{a_q},
\quad \forall q\in\mathcal{Q},
\label{eq:aq_unc}
\\
\mathbf A_q &= \hat{\mathbf A}_q + \Delta \mathbf A_q,
\qquad
\|\Delta\mathbf{A}_q\|_F \le \varepsilon_{A_q},
\quad \forall q\in\mathcal{Q},
\label{eq:Aq_unc}
\end{align}
where $\hat{\mathbf{h}}_k$, $\hat{\mathbf{a}}_q$, and $\hat{\mathbf{A}}_q$ are the channel estimates available at the BS, while
$\Delta\mathbf{h}_k$, $\Delta\mathbf{a}_q$, and $\Delta\mathbf{A}_q$ represent unknown estimation errors. Although $\mathbf A_q=\mathbf a_q\mathbf a_q^H$ in the nominal channel model, the forward channel estimate $\hat{\mathbf a}_q$ and the effective round-trip channel estimate $\hat{\mathbf A}_q$ are treated as separately estimated quantities to allow a mathematically tractable robust optimization.

\section{Robust Max--Min Sensing SINR Problem Formulation}
\label{sec_formulation}
In this section, we formulate the robust max–min sensing optimization problem for the proposed SIM-assisted covert ISAC system. The design is carried out under practical constraints including communication users QoS, total transmit power budget, and stringent covertness constraints that limit the detectability of covert transmissions at potential eavesdropping targets. Furthermore, to account for channel uncertainty, bounded CSI is incorporated, leading to a worst-case robust optimization framework. 

Our objective is to jointly design the transmit communication beamformers $\{\mathbf{w}_k\}_{k\in\mathcal K}$, radar covariance matrix $\mathbf{R}_0 \succeq \mathbf{0}$, receive sensing beamformers $\{\mathbf{u}_q\}_{q\in\mathcal Q}$, and SIM wave-domain beamforming matrices $\{\boldsymbol{\Theta}^l\}_{l\in\mathcal L}$ to maximize the minimum target sensing SINR. The resulting max--min sensing SINR problem is
\begin{subequations}
\begin{align}
\max_{\substack{\{\mathbf{w}_k\},\, \mathbf{R}_0, \\
\{\mathbf{u}_q\},\, \{\boldsymbol{\Theta}^l\}}}
 & \min_{q\in\mathcal{Q}} \min_{\|\Delta\mathbf{A}_q\|_F \le \varepsilon_{A_q},\forall q\in\mathcal{Q}} \gamma_q \label{Eq:Op1-a}\\
\text{s.t.} \quad
& \gamma_k \ge \Gamma_k,
\|\Delta\mathbf{h}_k\|_2 \le \varepsilon_{h_k},~ \forall k\in\mathcal{K}, \label{Eq:Op1-b}\\
& \sum_{k\in\mathcal K} \|\mathbf{w}_k\|^2
+ \operatorname{Tr}(\mathbf{R}_0)
\le P_{\max}, \label{Eq:Op1-c}\\
& T \, \nu\!\left(
\frac{\delta_{1,q} - \delta_{0,q}}{\delta_{0,q}}
\right)
\le 2\epsilon^2, \|\Delta\mathbf{a}_q\|_2 \le \varepsilon_{a_q},~ \forall q \in \mathcal Q, \label{Eq:Op1-d}\\
& \boldsymbol{\Theta}^l
= \operatorname{diag}\left(
e^{j\theta_{1}^l},\dots,e^{j\theta_{M}^l}
\right), |e^{j\theta_{m}^l}| = 1,
~ \forall l, \label{Eq:Op1-e}\\
& \mathbf{R}_0 \succeq 0 \label{Eq:Op1-f}.
\end{align}
\label{Eq:Op1}
\end{subequations}

The constraints~\eqref{Eq:Op1-b} ensure a minimum quality-of-service for each covert user, where $\Gamma_k = 2^{R_k^{\text{min}}}-1$ and $R_k^{\text{min}}$ denotes the minimum rate requirement for the $k$-th covert user. These constraints couple the communication beamformers $\{\mathbf{w}_k\}$, the radar covariance $\mathbf{R}_0$, and the SIM matrix $\mathbf{G}$. Since increasing the sensing power, through $\mathbf{R}_0$, improves the sensing energy but degrades the user SINR due to additional interference, the total transmit power constraint~\eqref{Eq:Op1-c} creates a fundamental tradeoff between communication beamforming power and sensing power allocation. Allocating more power to $\mathbf{R}_0$ improves sensing and covertness but reduces the available power for user beamforming. The KL-divergence constraint~\eqref{Eq:Op1-d} limits the detection capability at each target. 

The formulated optimization problem is highly non-convex due to the fractional SINR expressions in both communication and sensing, coupling between SIM phase shift matrices in the SIM transfer matrix $\mathbf{G}$, the unit-modulus constraints $|e^{j\theta_{m}^l}| = 1$, and the nonlinear KL-divergence covertness constraint. Before we dive into the problem solution, we first need to reformulate the problem into a more tractable form.

First, we recast the Max-Min sensing SINR problem by introducing the auxiliary variable $\varrho$ such that $\gamma_q \ge \varrho,~\forall q\in\mathcal{Q}, \|\Delta A_q\|_F\le\varepsilon_{A_q}$. Hence, problem~\eqref{Eq:Op1} can be expressed as
\begin{subequations}
\begin{align}
\max_{\substack{\{\mathbf{w}_k\},\, \mathbf{R}_0, \varrho \\
\{\mathbf{u}_q\},\, \{\boldsymbol{\Theta}^l\}}}
\quad & \varrho \label{Eq:Op2-a}\\
\text{s.t.} \quad
& \gamma_q \ge \varrho,\|\Delta \mathbf A_q\|_F\le\varepsilon_{A_q},~\forall q\in\mathcal{Q}, \label{Eq:Op2-b}\\
& \eqref{Eq:Op1-b} - \eqref{Eq:Op1-f}\label{Eq:Op2-c}.
\end{align}
\label{Eq:Op2}
\end{subequations}

Then, we have to convert the semi-infinite quadratic constraints induced by bounded CSI into LMIs.

\begin{lem}
S-procedure for a quadratic constraint~\cite{boyd1994linear}

Define the quadratic functions in the variable $\mathbf{z}\in \mathbb{C}^{n \times 1}$:
\begin{equation}
f(\mathbf z)=\mathbf z^H \mathbf A \mathbf z + 2\Re\{\mathbf b^H \mathbf z\} + c
\end{equation}
and
\begin{equation}
g(\mathbf z)=\mathbf z^H \mathbf C \mathbf z + 2\Re\{\mathbf d^H \mathbf z\} + e,
\end{equation}
where $\mathbf A,\mathbf C\in\mathbb C^{n\times n}$ are Hermitian, $\mathbf b,\mathbf d\in\mathbb C^{n\times 1}$, and $c,e\in\mathbb R$. The condition $\left(g(\mathbf z)\le 0 \;\Longrightarrow\; f(\mathbf z)\le 0\right)$ holds if and only if there exists a scalar $\mu\ge 0$ such that
\begin{equation}
\mu
\begin{bmatrix}
\mathbf C & \mathbf d\\
\mathbf d^H & e
\end{bmatrix}
-
\begin{bmatrix}
\mathbf A & \mathbf b\\
\mathbf b^H & c
\end{bmatrix}
\succeq \mathbf 0.
\end{equation}
\label{lem1}
\end{lem}

In the following, we apply Lemma~\ref{lem1} to constraints~\eqref{Eq:Op1-b},~\eqref{Eq:Op1-d}, and~\eqref{Eq:Op2-b}. By substituting $\mathbf h_k = \hat{\mathbf{h}}_k + \Delta \mathbf{h}_k$ into~\eqref{Eq:Op1-b}, $\gamma_k \ge \Gamma_k$ is given by
\begin{equation}
(\hat{\mathbf{h}}_k+\Delta\mathbf{h}_k)^H
\mathbf{Q}_k
(\hat{\mathbf{h}}_k+\Delta \mathbf{h}_k)
-
\Gamma_k \sigma_k^2
\ge 0,
\quad
\forall \|\Delta \mathbf{h}_k\|_2 \le \varepsilon_{h_k},
\label{eq:robust_sinr_quad2}
\end{equation}
where $\mathbf{Q}_k \triangleq \mathbf{G} \left(\mathbf{W}_k - \Gamma_k \sum_{i\in\mathcal K,\;i\neq k}\mathbf{W}_i - \Gamma_k \mathbf{R}_0 \right) \mathbf{G}^H$ and $\mathbf{W}_i = \mathbf{w}_i \mathbf{w}_i^H,~\forall i\in\mathcal{K}$. Expanding the quadratic term with respect to $\Delta \mathbf h_k$ yields
\small
\begin{dmath}
\Delta\mathbf h_k^H \mathbf Q_k \Delta\mathbf h_k
+
2\Re\left\{
\Delta\mathbf h_k^H \mathbf Q_k \hat{\mathbf h}_k
\right\}
+
\hat{\mathbf{h}}_k^H \mathbf Q_k \hat{\mathbf{h}}_k
-
\Gamma_k \sigma_k^2
\ge 0.
\label{eq:robust_sinr_quad3}
\end{dmath}
\normalsize

Since the uncertainty set $\|\Delta \mathbf h_k\|_2 \le \varepsilon_{h_k}$ is equivalently expressed as $\Delta\mathbf h_k^H \Delta\mathbf h_k - \varepsilon_{h_k}^2 \le 0$, the constraint in \eqref{eq:robust_sinr_quad3} holds if and only if there exists a scalar $\mu_k\ge 0$ such that
\begin{equation}
\begin{bmatrix}
\mu_k \mathbf I + \mathbf Q_k & \mathbf Q_k \hat{\mathbf{h}}_k\\
\hat{\mathbf{h}}_k^H \mathbf Q_k &
\hat{\mathbf{h}}_k^H \mathbf Q_k \hat{\mathbf{h}}_k - \Gamma_k \sigma_k^2 - \mu_k \varepsilon_{h_k}^2
\end{bmatrix}
\succeq \mathbf 0.
\label{Eq:robust_sinr_lmi}
\end{equation}

Therefore, the original robust SINR requirement under norm-bounded channel uncertainty is equivalently transformed into the LMI in \eqref{Eq:robust_sinr_lmi}, which is convex with respect to the variables $\mathbf W_i$, $\mathbf{R}_0$, and the auxiliary scalar $\mu_k$.

Regarding the robust covert constraint~\eqref{Eq:Op1-d}, it can be equivalently expressed as
\begin{equation}
\sum_{k\in\mathcal K}|\mathbf f_{1,q}\mathbf w_k|^2
-\varpi \mathbf{f}_{1,q}\mathbf{R}_0\mathbf{f}_{1,q}^H
\le \varpi \sigma_q^2,
\end{equation}
where $\varpi = \nu^{-1}\left(\frac{2\epsilon^2}{T}\right)$. Since $\mathbf f_{1,q}=\mathbf a_q^H\mathbf G$, we have $|\mathbf f_{1,q}\mathbf w_k|^2 = \mathbf a_q^H \mathbf G \mathbf W_k \mathbf G^H \mathbf a_q$
and $\mathbf{f}_{1,q}\mathbf{R}_0\mathbf{f}_{1,q}^H = \mathbf a_q^H \mathbf G \mathbf R_0 \mathbf G^H \mathbf a_q$. Therefore, constraint~\eqref{Eq:Op1-d} can be written as
\begin{equation}
\mathbf a_q^H \mathbf \Xi \mathbf a_q
\le \varpi \sigma_q^2,
\end{equation}
where $\mathbf \Xi \triangleq \mathbf G\left(\sum_{k\in\mathcal K}\mathbf W_k-\varpi \mathbf R_0\right)\mathbf G^H$.
By substituting
$\mathbf a_q=\hat{\mathbf a}_q+\Delta\mathbf a_q$ and following the same S-procedure as previously done, the above semi-infinite constraint holds if there exists an auxiliary variable $\mu_q\ge 0$ such that
\begin{equation}
\begin{bmatrix}
\mu_q \mathbf I-\mathbf \Xi & -\mathbf \Xi \hat{\mathbf a}_q\\
-\hat{\mathbf a}_q^H \mathbf \Xi &
\varpi \sigma_q^2-\hat{\mathbf a}_q^H \mathbf \Xi \hat{\mathbf a}_q-\mu_q \varepsilon_{a_q}^2
\end{bmatrix}
\succeq \mathbf 0. \label{Eq:robust_covert_lmi}
\end{equation}

Hence, the original robust covertness constraint is equivalently transformed into the above LMI.

To handle the semi-infinite constraint~\eqref{Eq:Op2-b} induced by the uncertainties in $\{\mathbf A_q\}_{q\in\mathcal Q}$, we follow a similar reformulation procedure as in~\cite{zhang2024robust}. Since $\mathbf F_q = \alpha_q \mathbf B^H \mathbf A_q \mathbf G$, we have
\begin{dmath}
\mathbf u_q^H \mathbf F_j \mathbf R_x \mathbf F_j^H \mathbf u_q
=
|\alpha_j|^2 \mathbf u_q^H \mathbf B^H \mathbf A_j \mathbf G \mathbf R_x \mathbf G^H  \mathbf A_j^H \mathbf B \mathbf u_q =
\operatorname{vec}(\mathbf A_j)^H \mathbf Y_{q,j} \operatorname{vec}(\mathbf A_j),
\end{dmath}
where $\mathbf Y_{q,j}
\triangleq
|\alpha_j|^2 \left( \left(\mathbf G \mathbf R_x \mathbf G^H \right)^T \otimes \mathbf B \mathbf U_q \mathbf B^H \right)$, and $\mathbf U_q = \mathbf u_q \mathbf u_q^H$.

Then, the constraint $\gamma_q \ge \varrho$ can be rewritten as
\begin{dmath}
\sum_{j\in\mathcal Q,\; j\neq q}
\operatorname{vec}(\mathbf A_j)^H \mathbf Y_{q,j} \operatorname{vec}(\mathbf A_j)
-
\frac{1}{\varrho}
\operatorname{vec}(\mathbf A_q)^H \mathbf Y_{q,q} \operatorname{vec}(\mathbf A_q)
+
\sigma_B^2 \operatorname{Tr}(\mathbf U_q)
\le 0. \label{Eq:sens_const}
\end{dmath}

To obtain a single quadratic uncertainty constraint, we define
$
\hat{\mathbf g}_A = \left[
\operatorname{vec}(\hat{\mathbf A}_1)^T, \dots,
\operatorname{vec}(\hat{\mathbf A}_{Q})^T
\right]^T\in \mathbb{C}^{\left(QM^2\times1\right)}$ and $
\Delta \mathbf g_A = \left[
\operatorname{vec}(\Delta \mathbf A_1)^T, \dots,
\operatorname{vec}(\Delta \mathbf A_{Q})^T
\right]^T \in \mathbb{C}^{\left(QM^2\times1\right)}$, where $\|\Delta \mathbf g_A\|_2^2 \le \sum_{q\in\mathcal Q} \varepsilon_{A_j}^2$. Then define the block diagonal matrix
\begin{equation}
\mathbf X_q =
\operatorname{diag}\left(
\mathbf Y_{q,1}, \ldots, \mathbf Y_{q,q-1},
\mathbf D_q,
\mathbf Y_{q,q+1}, \ldots, \mathbf Y_{q,|\mathcal Q|}
\right),
\end{equation}
where $\mathbf D_q = -\frac{1}{\varrho} \mathbf Y_{q,q}$.

Thus, the robust constraint is recast as
\begin{dmath}
\Delta \mathbf g_A^H \mathbf X_q \Delta \mathbf g_A
+ 2 \Re \left\{ (\mathbf X_q \hat{\mathbf g}_A)^H \Delta \mathbf g_A \right\}
+ \hat{\mathbf g}_A^H \mathbf X_q \hat{\mathbf g}_A
+ \sigma_B^2 \operatorname{Tr}(\mathbf U_q)
\le 0,
\quad \forall \Delta \mathbf g_A.
\end{dmath}

By applying the S-procedure in Lemma~\ref{lem1}, the above constraint holds if there exists $\mu_{A_q} \ge 0$ such that
\small
\begin{dmath}
\begin{bmatrix}
\mu_{A_q} \mathbf I - \mathbf X_q
&
-\mathbf X_q \hat{\mathbf g}_A
\\
-\hat{\mathbf g}_A^H \mathbf X_q
&
-\hat{\mathbf g}_A^H \mathbf X_q \hat{\mathbf g}_A
- \sigma_B^2 \operatorname{Tr}(\mathbf U_q)
- \mu_{A_q} \sum_{j\in\mathcal Q} \varepsilon_{A_j}^2
\end{bmatrix}
\succeq \mathbf 0. \label{Eq:robust_sensing_lmi}
\end{dmath}
\normalsize

Therefore, the original semi-infinite sensing constraint can be sufficiently replaced with the above LMI. Now, our optimization problem can be expressed as follows
\begin{subequations}
\begin{align}
\max_{\substack{\{\mathbf{W}_k\},\, \mathbf{R}_0, \varrho \\
\{\mathbf{U}_q\},\, \{\boldsymbol{\Theta}^l\}}}
\quad & \varrho \label{Eq:Op3-a}\\
\text{s.t.} \quad
& \eqref{Eq:Op1-c}, \eqref{Eq:Op1-e} - \eqref{Eq:Op1-f}, \label{Eq:Op3-b}\\
& \eqref{Eq:robust_sinr_lmi}, \eqref{Eq:robust_covert_lmi}, \eqref{Eq:robust_sensing_lmi} \label{Eq:Op3-c}.
\end{align}
\label{Eq:Op3}
\end{subequations}
\vspace{-0.7cm}
\section{Alternating Optimization-based Robust Max--Min Sensing SINR Problem Solution}
\label{sec_opti}
Due to the coupled dependence of $\{\mathbf W_k\}$, $\mathbf R_0$, $\{\mathbf U_q\}$, and $\{\boldsymbol\Theta^l\}$ in \eqref{Eq:robust_sinr_lmi}, \eqref{Eq:robust_covert_lmi}, and \eqref{Eq:robust_sensing_lmi}, problem \eqref{Eq:Op3} is non-convex and difficult to solve directly. To tackle this issue, we adopt an AO framework by partitioning the optimization variables into three blocks, namely the receive variables $\{\mathbf U_q\}$, the transmit variables $\{\mathbf W_k\},\mathbf R_0$,  and the SIM phase matrices $\{\boldsymbol\Theta^l\}$.

\subsection{Receive Beamforming Vectors Optimization Sub-problem}
For fixed $\{\mathbf W_k\},\mathbf R_0$, and $\{\boldsymbol\Theta^l\}$, the receive-side matrices $\{\mathbf U_q\}$ only appear in $\mathbf X_q$ through affine mappings. Therefore, the receive-side sub-problem can be expressed as
\begin{subequations}
\begin{align}
\max_{\{\mathbf U_q\},\,\varrho}\quad & \varrho \\
\text{s.t.}\quad
& \mathbf U_q\succeq 0, \operatorname{Tr}(\mathbf U_q)=1,\quad \forall q\in\mathcal Q, \label{Eq:50-b}\\
& \operatorname{Rank}(\mathbf U_q)=1,\quad \forall q\in\mathcal Q,\\
& \eqref{Eq:robust_sensing_lmi}
\end{align}
\label{Eq:rec_beam}
\end{subequations}

Problem~\eqref{Eq:rec_beam} is still non-convex due to the non-convex rank-one set of constraints $\operatorname{Rank}(\mathbf U_q)=1$. Moreover, it is worth noting that the matrix $\mathbf X_q$ contains the term $\mathbf D_q=-\frac{1}{\varrho}\mathbf Y_{q,q}$, which makes the problem non-convex in $\left\{\varrho,\mathbf U_q\right\}$. In order to handle the non-convex rank-one constraints $\operatorname{Rank}(\mathbf U_q)=1$, we use the following equivalence for any Hermitian positive semidefinite matrix $\mathbf U_q$
\begin{equation}
\operatorname{Rank}(\mathbf U_q)=1
\iff
\operatorname{Tr}(\mathbf U_q)-\lambda_{\max}(\mathbf U_q)=0.
\end{equation}

This follows from the fact that $\operatorname{Tr}(\mathbf U_q)$ equals the sum of all eigenvalues of $\mathbf U_q$, whereas $\lambda_{\max}(\mathbf U_q)$ is its largest eigenvalue. Since $\mathbf U_q\succeq \mathbf 0$, $\operatorname{Tr}(\mathbf U_q)\ge \lambda_{\max}(\mathbf U_q)$ holds with equality if and only if only one eigenvalue is nonzero, i.e., when $\mathbf U_q$ is rank one. Therefore, the rank-one constraints can be replaced by introducing the penalty term $\Big[\zeta_1 \sum_{q\in\mathcal Q}\left(\operatorname{Tr}(\mathbf U_q)-\lambda_{\max}(\mathbf U_q)\right)\Big]$ into the objective function, where $\zeta_1>0$ is a penalty coefficient. Then, we adopt the successive convex approximation (SCA) to deal with the part $-\lambda_{\max}(\mathbf U_q)$. In the $r$-th iteration, let $\mathbf v_q^{(r)}$ denote the dominant eigenvector of the current point $\mathbf U_q^{(r)}$. By the Rayleigh quotient inequality,
\begin{equation}
\lambda_{\max}(\mathbf U_q)\ge (\mathbf v_q^{(r)})^H \mathbf U_q \mathbf v_q^{(r)},
\end{equation}
where $\lambda_{\max}(\mathbf U_q)$ is replaced by its affine lower bound. To remove the non-convexity caused by the reciprocal term $\frac{1}{\varrho}$ in $\mathbf D_q$, we introduce a new scalar auxiliary variable $\bar{\varrho} \triangleq \frac{1}{\varrho}, ~ \bar{\varrho}>0$ such that $\mathbf D_q = -\bar{\varrho} \mathbf Y_{q,q}$. Additionally, we adopt a quasi-convex reformulation based on fixing $\bar{\varrho}$ and applying a bisection search. Accordingly, the original problem can be solved by searching for the minimum $\bar{\varrho}$ such that the resulting SDP is feasible. At each iteration, a candidate $\bar{\varrho}$ is tested by solving the corresponding SDP feasibility problem. If the problem is feasible, $\bar{\varrho}$ is a valid upper bound; otherwise, it is a lower bound. Then, the interval is iteratively refined until convergence. Therefore, the optimal value $\varrho^\star$ is obtained as the smallest $\bar{\varrho}$ for which the SDP is feasible, while the corresponding $\{\mathbf{U}_q\}$ are obtained from the final SDP solution. Accordingly, at the candidate $\bar{\varrho}$, the resulting SDP inner problem at the $r$-th iteration is given as
\begin{subequations}
\begin{align}
\min_{\{\mathbf U_q\}}\quad & \bar{\varrho}+
\zeta_1\sum_{q\in\mathcal Q}
\left(
\operatorname{Tr}(\mathbf U_q)-(\mathbf v_q^{(r)})^H\mathbf U_q\mathbf v_q^{(r)}
\right)
\\
\text{s.t.}\quad
& \eqref{Eq:robust_sensing_lmi}, \eqref{Eq:50-b}.
\end{align}
\label{eq:rx_sca_obj}
\end{subequations}

The problem \eqref{eq:rx_sca_obj} is a convex SDP and can be efficiently solved using standard convex optimization solvers~\cite{grant2009cvx}. 

Since the robust sensing constraint~\eqref{Eq:robust_sensing_lmi} involves the matrix $\mathbf X_q$, whose direct construction requires high-dimensional Kronecker products, the direct implementation of the robust LMI sensing constraint adds extremely high computational complexity to the problem. In particular, recall that
$\mathbf{Y}_{q,j} \triangleq |\alpha_j|^2
\left[
\left(\mathbf{G}\mathbf{R}_x\mathbf{G}^{H}\right)^{T}
\otimes
\mathbf{B}\mathbf{U}_q\mathbf{B}^{H}
\right]$
and that the matrix $\mathbf{X}_q$ is constructed by stacking the matrices $\mathbf{Y}_{q,j}$ in a block-diagonal manner. Hence, as $\mathbf{G}\mathbf{R}_x\mathbf{G}^{H}\in\mathbb{C}^{M\times M}$ and $\mathbf{B}\mathbf{U}_q\mathbf{B}^{H}\in\mathbb{C}^{M\times M}$, each matrix $\mathbf{Y}_{q,j}$ has dimension $M^2\times M^2$, whereas $\mathbf{X}_q$ has dimension $QM^2\times QM^2$. Consequently, the corresponding S-procedure LMI has dimension $(QM^2+1)\times(QM^2+1)$. This excessive growth with the number of meta-atoms is computationally demanding in terms of both memory and processing complexity.

To significantly reduce the computational complexity while preserving the exact structure of the robust LMI, we present the following Proposition.

\begin{pro}
\label{thm:dimension_reduction}
The high-dimensional robust sensing LMI constraint~\eqref{Eq:robust_sensing_lmi} can be represented exactly in a lower-dimensional space as
\small
\begin{dmath}
\begin{bmatrix}
\mu_{A_q}\mathbf I-\widetilde{\mathbf K}_q
&
-\widetilde{\mathbf K}_q\widetilde{\mathbf g}_A
\\
-\widetilde{\mathbf g}_A^H\widetilde{\mathbf K}_q
&
-\widetilde{\mathbf g}_A^H\widetilde{\mathbf K}_q\widetilde{\mathbf g}_A
-\sigma_B^2\operatorname{Tr}(\mathbf U_q)
-\mu_{A_q}\displaystyle\sum_{j\in\mathcal Q}\varepsilon_{A_j}^2
\end{bmatrix}
\succeq \mathbf 0. \label{Eq:red_LMI}
\end{dmath}
\normalsize
where $\widetilde{\mathbf K}_q = \mathbf R_C \mathbf K_q \mathbf R_C^H$, $\widetilde{\mathbf g}_A = \mathbf U_C^H \hat{\mathbf g}_A$, and $\mathbf U_C, \mathbf R_C$ are obtained from the compact QR factorization $\mathbf C = \mathbf U_C \mathbf R_C$ of $\mathbf C = \operatorname{diag}\left(\mathbf C_1,\ldots,\mathbf C_Q\right)$ with $\mathbf C_j = \vert{}\alpha_j\vert{} \left(\mathbf U_G^* \otimes \mathbf B\right)$.
\end{pro}
Proof: See the Appendix.

The dimension reduction itself is exact and does not introduce additional approximation into the robust LMI. In particular, instead of operating in the original space whose dimension grows proportionally to $QM^2$, the reduced LMI operates in a space whose dimension grows proportionally to $Qr_GN_r$, where typically $r_G\leq N_t\ll M$. This substantially reduces the computational and memory requirements.

\subsection{Transmit Beamforming and Covariance Matrix Optimization Sub-problem}
For fixed $\{\mathbf U_q\},\forall q\in\mathcal{Q}$ and $\{\boldsymbol\Theta^l\}, \forall l\in\mathcal{L}$, the matrices $\mathbf Q_k$,  $\mathbf \Xi$, and $\mathbf X_q$ become affine functions of $\{\mathbf W_k\},\forall k \in \mathcal{K}$ and $\mathbf R_0$. Hence, the resulting transmit beamforming and covariance optimization sub-problem is
\begin{subequations}
\begin{align}
\max_{\{\mathbf{W}_k\},\, \mathbf{R}_0, \varrho}
\quad & \varrho \label{Eq:Op4-a}\\
\text{s.t.} \quad
& \sum_{k\in\mathcal K} \operatorname{Tr}(\mathbf{W}_k)
+ \operatorname{Tr}(\mathbf{R}_0)
\le P_{\max}, \label{Eq:Op4-b}\\
& \mathbf W_k\succeq0,~\forall k \in \mathcal{K}, \label{Eq:Op4-c} \\ 
& \text{Rank}\left(\mathbf W_k\right)=1,~\forall k \in \mathcal{K}, \label{Eq:Op4-d} \\
& \mathbf R_0\succeq0, \label{Eq:Op4-e}\\
& \eqref{Eq:robust_sinr_lmi}, \eqref{Eq:robust_covert_lmi}, \eqref{Eq:robust_sensing_lmi} \label{Eq:Op4-f}.
\end{align}
\label{Eq:Op4}
\end{subequations}

Problem~\eqref{Eq:Op4} is still non-convex due to the rank-one constraint of the transmit beamforming matrices and due to the coupling between the variables $\varrho$ and $\mathbf W_k$ within the matrix $\mathbf{X}_q$ through the term $\mathbf{D}_q = -\frac{1}{\varrho}\mathbf{Y}_{q,q}$. To handle the non-convex rank-one constraints $\operatorname{Rank}(\mathbf W_k)=1$, we use the same procedures as in the receive beamforming matrices. Therefore, the rank-one constraints can be replaced by introducing the penalty term $\Big[\zeta_2 \sum_{k\in\mathcal K}\left(\operatorname{Tr}(\mathbf W_k)-(\mathbf v_k^{(r)})^H \mathbf W_k \mathbf v_k^{(r)}\right)\Big]$ into the objective function, where $\zeta_2>0$ is a penalty coefficient. In addition, we introduce the auxiliary variable $\bar{\varrho} \triangleq \frac{1}{\varrho}, ~ \bar{\varrho}>0$ such that $\mathbf D_q = -\bar{\varrho} \mathbf Y_{q,q}$ and apply a bisection search over $\bar{\varrho}$ as in the receive beamforming sub-problem. As a result, the convex surrogate problem at the $r$-th iteration is given by
\begin{subequations}
\begin{align}
\min_{\{\mathbf W_k\},\,\mathbf R_0}
\quad &
\bar{\varrho}
+
\zeta_2 \sum_{k\in\mathcal K}
\left(
\operatorname{Tr}(\mathbf W_k)
-
(\mathbf v_k^{(r)})^H \mathbf W_k \mathbf v_k^{(r)}
\right)
\label{Eq:Op5-a}\\
\text{s.t.}\quad
& \eqref{Eq:Op4-b},\eqref{Eq:Op4-c},\eqref{Eq:Op4-e},\eqref{Eq:robust_sinr_lmi}, \eqref{Eq:robust_covert_lmi}, \eqref{Eq:robust_sensing_lmi} \label{Eq:Op5-b}.
\end{align}
\label{Eq:Op5}
\end{subequations}
For fixed $\{\mathbf v_k^{(r)}\}$, the objective in \eqref{Eq:Op5-a} is affine in the optimization variables. Hence, \eqref{Eq:Op5} is a convex semidefinite program and can be efficiently solved by standard convex optimization solvers~\cite{grant2009cvx}. The above procedure is iterated until convergence. Then, the transmit beamformer can be recovered by eigenvalue decomposition.

Similar to the receive beamforming sub-problem, the direct implementation of the robust sensing constraint~\eqref{Eq:robust_sensing_lmi} involves an LMI of dimension $(QM^2+1)\times(QM^2+1)$. A similar reduced-dimensional manipulation is employed, where the receive beamforming matrices $\{\mathbf U_q\}$ are fixed, while $\mathbf R_x=\sum_{k\in\mathcal K}\mathbf W_k+\mathbf R_0$ is an optimization variable. We therefore exploit the same column-space reduction principle while adapting the factorization to the structure of the transmit beamforming sub-problem. In particular, since the fixed receive covariance matrix satisfies $\mathbf U_q=\mathbf u_q\mathbf u_q^H$, we define $\mathbf b_q=\mathbf B\mathbf u_q$, such that $\mathbf B\mathbf U_q\mathbf B^H=\mathbf b_q\mathbf b_q^H$ is rank one. Moreover, using the compact SVD of $\mathbf G$, we express $\mathbf G\mathbf R_x\mathbf G^H=\mathbf U_G\mathbf T\mathbf U_G^H$, where $\mathbf T\in\mathbb C^{r_G\times r_G}$ remains affine in $\mathbf R_x$. Accordingly, each high-dimensional block can be factorized as $\mathbf Y_{q,j}=\mathbf C_{q,j}\mathbf T^T\mathbf C_{q,j}^H$, where $\mathbf C_{q,j}=|\alpha_j|(\mathbf U_G^*\otimes\mathbf b_q)$ is fixed. By subsequently applying a compact QR factorization to the block-diagonal matrix constructed from $\{\mathbf C_{q,j}\}_{j\in\mathcal Q}$, the original robust sensing LMI is represented exactly in a reduced subspace of dimension $Qr_G+1$, instead of $QM^2+1$. Hence, although both beamforming sub-problems employ the same underlying column-space reduction principle, the fixed rank-one structure of $\mathbf U_q$ in the transmit beamforming sub-problem yields the further dimensional reduction from $Qr_GN_r+1$ in~\eqref{Eq:red_LMI} to $Qr_G+1$.

\subsection{Layer-by-Layer SIM Phase Shift Optimization Sub-problem}
For fixed $\{\mathbf W_k\},\mathbf R_0$, and $\{\mathbf U_q\}$, we adopt a layer-by-layer framework to optimize the SIM phase shift matrices $\{\boldsymbol\Theta^l\}$ and optimize one layer at a time while fixing all the remaining layers. Since the target SINR variable $\varrho$ appears in the matrix block $\mathbf D_q=-\frac{1}{\varrho}\mathbf Y_{q,q}$, we introduce the auxiliary variable $\bar{\varrho} \triangleq \frac{1}{\varrho}, ~ \bar{\varrho}>0$ such that $\mathbf D_q = -\bar{\varrho} \mathbf Y_{q,q}$ and apply a bisection search over $\bar{\varrho}$ as in the previous sub-problems. At the $l$-th SIM layer, the forward and reverse wave-domain transfer matrices can be written as $\mathbf G=\mathbf C_L^l\boldsymbol{\Theta}^l\mathbf C_R^l$ and $
\mathbf B=\mathbf C_L^l\boldsymbol{\Theta}^l\widetilde{\mathbf C}_R^l$, respectively, 
where the matrices $\mathbf C_L^l,\mathbf C_R^l$, and $\widetilde{\mathbf C}_R^l$ are fixed with respect to the $l$-th layer, which can be expressed as
\begin{equation}
        \mathbf{C}_L^l = 
\begin{cases}
\mathbf{\Theta}^L \mathbf{\Omega}^L \dots \mathbf{\Theta}^{l+1} \mathbf{\Omega}^{l+1}, & \text{if } l \neq L, \\
\mathbf{I}_M, & \text{if } l = L,
\end{cases} \label{Eq:31}
    \end{equation}
    \begin{equation}
        \mathbf{C}_R^l = 
\begin{cases}
\mathbf{\Omega}^l\mathbf{\Theta}^{l-1}\mathbf{\Omega}^{l-1} \dots \mathbf{\Theta}^{1} \mathbf{\Omega}_t^{1}, & \text{if } l \neq 1, \\
\mathbf{\Omega}_t^{1}, & \text{if } l = 1,
\end{cases} \label{Eq:32}
    \end{equation}
and
    \begin{equation}
        \widetilde{\mathbf C}_R^l = 
\begin{cases}
\mathbf{\Omega}^l\mathbf{\Theta}^{l-1}\mathbf{\Omega}^{l-1} \dots \mathbf{\Theta}^{1} \mathbf{\Omega}_r^{1}, & \text{if } l \neq 1, \\
\mathbf{\Omega}_r^{1}, & \text{if } l = 1,
\end{cases} \label{Eq:33}
    \end{equation}
where \( \mathbf{\Omega}_r^1 \in \mathbb{C}^{M \times N_r} \) models the propagation coefficient matrix between the receive antenna array and the first SIM layer, which can be derived from~\eqref{Eq:2} by replacing 
\( r^l_{m,\tilde{m}} \) and \( \chi^l_{m,\tilde{m}} \) 
with \( r^1_{m,n_r} \) and \( \chi^1_{m,n_r} \), respectively, 
for all \( m \in \mathcal{M} \) and \( n_r \in \mathcal{N}_r \).
\subsubsection{LMI-Based Robust Sensing Constraint Analysis}
\leavevmode\par
Define $\pmb{\phi}_l= \left[e^{j\theta_1^l},\dots,e^{j\theta_M^l}\right]^T$ such that $\mathbf \Phi_l=\pmb{\phi}_l\pmb{\phi}_l^H$, $\mathbf \Phi_l \succeq \mathbf 0$, $\operatorname{diag}(\mathbf \Phi_l)=\mathbf 1$, and $\operatorname{Rank}(\mathbf \Phi_l)=1$. Then, the term $\mathbf G\mathbf R_x\mathbf G^H$ in $\mathbf Y_{q,j},~\forall q,j \in \mathcal{Q}$ can be written as
\begin{dmath}
\mathbf{Z}_G = \mathbf G\mathbf R_x\mathbf G^H
=
\mathbf C_L^l\boldsymbol{\Theta}^l\mathbf C_R^l \mathbf R_x \left(\mathbf C_R^l\right)^H \left(\boldsymbol{\Theta}^l\right)^H \left(\mathbf C_L^l\right)^H = \mathbf C_L^l \operatorname{diag}(\pmb{\phi}_l) \mathbf C_R^l \mathbf R_x \left(\mathbf C_R^l\right)^H \operatorname{diag}(\pmb{\phi}_l)^H \left(\mathbf C_L^l\right)^H = \mathbf C_L^l \left( \left( \mathbf C_R^l \mathbf R_x \left(\mathbf C_R^l\right)^H \right) \odot \mathbf \Phi_l \right) \left(\mathbf C_L^l\right)^H,
\end{dmath}
where $\odot$ is the element-wise product or the Hadamard product. Similarly, the reverse term $\mathbf B \mathbf U_q \mathbf B^H$ can be written as
\begin{equation}
\mathbf{Z}_{B,q} = \mathbf B\mathbf U_q\mathbf B^H = \mathbf C_L^l \left( \left( \widetilde{\mathbf C}_R^l \mathbf U_q \left(\widetilde{\mathbf C}_R^l\right)^H \right) \odot \mathbf \Phi_l \right) \left(\mathbf C_L^l\right)^H.
\end{equation}

For fixed transmit and receive beamformers, both $\mathbf Z_G$ and $\mathbf Z_{B,q}$ are affine in $\mathbf \Phi_l$, whereas, $\mathbf Y_{q,j}$ is bilinear in these matrices. To obtain a tractable approximation, at the $r$-th iteration we use the first-order SCA approximation to linearize $\mathbf Y_{q,j}$ around the current point $\mathbf \Phi_l^{(r)}$ as follows
\begin{equation}
\mathbf Y_{q,j}^{(r)} = |\alpha_j|^2 \Big(\mathbf Z_G^T\otimes \mathbf Z_{B,q}^{(r)} + (\mathbf Z_G^{(r)})^T\otimes \mathbf Z_{B,q} - (\mathbf Z_G^{(r)})^T\otimes \mathbf Z_{B,q}^{(r)}
\Big).
\end{equation}

Then, we substitute the approximation $\mathbf Y_{q,j}^{(r)}$ into $\mathbf{X}_q$ to get $\mathbf{X}_q^{(r)}$, which will be used to approximate the LMI in~\eqref{Eq:robust_sensing_lmi}. 

In similar ways, $\mathbf Q_k$ in~\eqref{Eq:robust_sinr_lmi} and $\mathbf \Xi$~\eqref{Eq:robust_covert_lmi} can be expressed as
\begin{equation}
\mathbf{Z}_{Q_k} = \mathbf G\mathbf S_{Q_k}\mathbf G^H = \mathbf C_L^l \left( \left( \mathbf C_R^l \mathbf S_{Q_k} \left(\mathbf C_R^l\right)^H \right) \odot \mathbf \Phi_l \right) \left(\mathbf C_L^l\right)^H,
\end{equation}
and
\begin{equation}
\mathbf{Z}_{\Xi} = \mathbf G\mathbf S_{\Xi}\mathbf G^H = \mathbf C_L^l \left( \left( \mathbf C_R^l \mathbf S_{\Xi} \left(\mathbf C_R^l\right)^H \right) \odot \mathbf \Phi_l \right) \left(\mathbf C_L^l\right)^H,
\end{equation}
respectively, where $\mathbf S_{Q_k} = \mathbf{W}_k - \Gamma_k \sum_{i\in\mathcal K,\;i\neq k}\mathbf{W}_i - \Gamma_k \mathbf{R}_0$ and $\mathbf S_{\Xi} = \sum_{k\in\mathcal K}\mathbf W_k-\varpi \mathbf R_0$.

As in previous sub-problems, to enforce the rank-one property of $\mathbf \Phi_l$, we add the penalty term $\Big[\zeta_3
\left(
\operatorname{Tr}(\mathbf \Phi_l)-(\pmb{\phi}_l^{(r)})^H \mathbf \Phi_l \pmb{\phi}_l^{(r)}
\right)\Big]$ to the objective function, where $\pmb{\phi}_l^{(r)}$ is the dominant eigenvector of $\mathbf \Phi_l^{(r)}$. Consequently, for a fixed $\bar{\varrho}$, the $l$-th layer phase update is obtained by
\begin{subequations}
\begin{align}
\min_{\mathbf \Phi_l}\quad
& \bar{\varrho} + \zeta_3
\left(
\operatorname{Tr}(\mathbf \Phi_l)-(\pmb{\phi}_l^{(r)})^H \mathbf \Phi_l \pmb{\phi}_l^{(r)}
\right) \\
\text{s.t.}\quad
& \mathbf \Phi_l \succeq \mathbf 0,~\operatorname{diag}(\mathbf \Phi_l)=\mathbf 1\\
 &\eqref{Eq:robust_sinr_lmi},~\eqref{Eq:robust_covert_lmi}, \text{and}~\eqref{Eq:robust_sensing_lmi}.
\end{align}
\label{Eq:phase_shift_prob}
\end{subequations}
\vspace{-0.5cm}

The problem~\eqref{Eq:phase_shift_prob} is a convex SDP problem that can be solved by standard convex optimization solvers~\cite{grant2009cvx}. The above procedure is repeated sequentially for all layers $l \in \mathcal{L}$. For a given trial value $\bar{\varrho}$, if~\eqref{Eq:phase_shift_prob} is feasible,
then $\bar{\varrho}$ is declared feasible; otherwise, it is infeasible. Therefore, the optimal $\bar{\varrho}$ is obtained by an outer bisection search.
\subsubsection{Low-Complexity Robust Sensing Constraint Analysis}
\leavevmode\par
As discussed previously, directly implementing the robust sensing LMI constraint,~\eqref{Eq:robust_sensing_lmi}, entails substantial computational complexity and memory overhead. Unfortunately, the exact dimension-reduction technique developed for the transmit and receive beamforming sub-problems cannot be applied to this sub-problem, as the optimization variables now appear on both sides of the Kronecker product in $\mathbf{Y}_{q,j}$. To address this challenge, we propose an alternative low-complexity solution specifically tailored to the SIM phase shift sub-problem in this section. The sensing constraint in~\eqref{Eq:sens_const} can be rewritten as
\begin{equation}
    \sum_{j\neq q} \psi_{q,j} - \bar{\varrho}\, \psi_{q,q} + \sigma_B^2 \operatorname{Tr}(\mathbf U_q) \le 0,
\quad \forall q\in\mathcal Q, \label{eq:def_psi_qj}
\end{equation}
where $\bar{\varrho} = \frac{1}{\varrho}$ and  $\psi_{q,j} =\operatorname{vec}(\mathbf A_j)^H \mathbf Y_{q,j} \operatorname{vec}(\mathbf A_j)\triangleq |\alpha_j|^2 \operatorname{Tr}\!\left( \mathbf Z_{B,q}\, \mathbf A_j\, \mathbf Z_G\, \mathbf A_j^H \right)$. Substituting $\mathbf A_j = \hat{\mathbf A}_j + \Delta \mathbf A_j,~ \forall j \in \mathcal{Q}$ into \eqref{eq:def_psi_qj} yields
\begin{multline}
\psi_{q,j}
=
|\alpha_j|^2
\Big( 
\operatorname{Tr}\!\left(
\mathbf Z_{B,q}\hat{\mathbf A}_j \mathbf Z_G \hat{\mathbf A}_j^H
\right) \\ +
2\Re\!\left\{
\operatorname{Tr}\!\left(
\mathbf Z_{B,q}\Delta \mathbf A_j \mathbf Z_G \hat{\mathbf A}_j^H
\right)
\right\} +
\operatorname{Tr}\!\left(
\mathbf Z_{B,q}\Delta \mathbf A_j \mathbf Z_G \Delta \mathbf A_j^H
\right)
\Big).
\label{eq:psi_expand}
\end{multline}

From the Cauchy--Schwarz inequality associated with the
Frobenius inner product~\cite[Sec.~1.1]{watrous2018theory}, for any two
complex matrices $\mathbf X$ and $\mathbf Y$ of compatible
dimensions, we have
\begin{equation}
\left|\operatorname{Tr}(\mathbf X^H\mathbf Y)\right|
\leq
\|\mathbf X\|_F\|\mathbf Y\|_F.
\label{eq:frob_cs}
\end{equation}

Consequently, under the bounded uncertainty model
$\|\Delta\mathbf A_j\|_F\leq\varepsilon_{A_j}$,
\begin{equation}
\max_{\|\Delta\mathbf A_j\|_F\leq\varepsilon_{A_j}}
\Re\left\{
\operatorname{Tr}(\Delta\mathbf A_j^H\mathbf Y)
\right\}
=
\varepsilon_{A_j}\|\mathbf Y\|_F.
\label{eq:frob_duality}
\end{equation}

Applying~\eqref{eq:frob_duality} to the cross term in \eqref{eq:psi_expand}, we obtain
\begin{dmath}
2\Re\!\left\{
\operatorname{Tr}\!\left(
\mathbf Z_{B,q}\Delta \mathbf A_j \mathbf Z_G \hat{\mathbf A}_j^H
\right)
\right\}
=
2\Re\!\left\{
\operatorname{Tr}\!\left(
\Delta \mathbf A_j \mathbf \Psi_{q,j}
\right)
\right\}
=
2\varepsilon_{A_j} \|\mathbf \Psi_{q,j}\|_F,
\end{dmath}
where $\mathbf \Psi_{q,j}
\triangleq
\mathbf Z_G \widehat{\mathbf A}_j^H \mathbf Z_{B,q}$.

Moreover, since $\mathbf Z_G\succeq\mathbf 0$ and
$\mathbf Z_{B,q}\succeq\mathbf 0$, using the standard
submultiplicative property of Schatten norms
\cite[Sec.~1.1]{watrous2018theory},
the quadratic uncertainty term satisfies
\begin{dmath}
\operatorname{Tr}\left(
\mathbf Z_{B,q}\Delta\mathbf A_j
\mathbf Z_G\Delta\mathbf A_j^H
\right)
=
\left\|
\mathbf Z_{B,q}^{1/2}
\Delta\mathbf A_j
\mathbf Z_G^{1/2}
\right\|_F^2
\leq
\|\mathbf Z_{B,q}\|_2 \|\Delta\mathbf A_j\|_F^2
\|\mathbf Z_G\|_2
\leq
\varepsilon_{A_j}^2
\|\mathbf Z_{B,q}\|_2
\|\mathbf Z_G\|_2.
\label{eq:quadratic_error_bound}
\end{dmath}

Therefore, a robust upper bound for the interference term ($j\neq q$) is
\begin{dmath}
\overline{\psi}_{q,j}
=
|\alpha_j|^2
\Big(
\operatorname{Tr}\!\left(
\mathbf Z_{B,q}\hat{\mathbf A}_j \mathbf Z_G \hat{\mathbf A}_j^H
\right)
+
2\varepsilon_{A_j}\eta_{q,j}
+
\varepsilon_{A_j}^2 \beta_q
\Big),
\label{eq:upper_interference}
\end{dmath}
while a robust lower bound for the desired term ($j=q$) is
\begin{dmath}
\underline{\psi}_{q,q}
=
|\alpha_q|^2
\Big(
\operatorname{Tr}\!\left(
\mathbf Z_{B,q}\hat{\mathbf A}_q \mathbf Z_G \hat{\mathbf A}_q^H
\right)
-
2\varepsilon_{A_q}\eta_{q,q}
-
\varepsilon_{A_q}^2 \beta_q
\Big),
\label{eq:lower_desired}
\end{dmath}
where $\eta_{q,j}$ and $\beta_q$ are defined as
\begin{equation}
\eta_{q,j}
\triangleq
\left\|
\mathbf \Psi_{q,j}
\right\|_F, \qquad
\beta_q
\triangleq
\|\mathbf Z_{B,q}\|_2 \|\mathbf Z_G\|_2.
\label{eq:def_eta_beta}
\end{equation}

Using~\eqref{eq:upper_interference} and~\eqref{eq:lower_desired}, a norm-bound-based sensing surrogate is
\begin{equation}
\sum_{j\neq q} \overline{\psi}_{q,j}
-
\bar{\varrho}\, \underline{\psi}_{q,q}
+
\sigma_B^2 \operatorname{Tr}(\mathbf U_q)
\le 0,
\qquad \forall q\in\mathcal Q.
\label{eq:robust_sensing_surrogate_exact}
\end{equation}

The bounds in \eqref{eq:upper_interference}--\eqref{eq:lower_desired} are still non-convex because both $\mathbf Z_G(\mathbf \Phi_l)$ and $\mathbf Z_{B,q}(\mathbf \Phi_l)$ depend on $\mathbf \Phi_l$, and their product appears in the nominal term. To obtain a convex sub-problem, we adopt an SCA step at the $r$-th iteration. Hence, the nominal term is approximated by the first-order affine expression

\small
\begin{multline}
\psi_{q,j}^{(r)}(\mathbf \Phi_l)
\triangleq 
\operatorname{Tr}\!\left(
\mathbf Z_{B,q}^{(r)} \widehat{\mathbf A}_j \mathbf Z_G(\mathbf \Phi_l)\widehat{\mathbf A}_j^H
\right) \\ +
\operatorname{Tr}\!\left(
\mathbf Z_{B,q}(\mathbf \Phi_l)\widehat{\mathbf A}_j \mathbf Z_G^{(r)} \widehat{\mathbf A}_j^H
\right) -
\operatorname{Tr}\!\left(
\mathbf Z_{B,q}^{(r)} \widehat{\mathbf A}_j \mathbf Z_G^{(r)} \widehat{\mathbf A}_j^H
\right),
\label{eq:psi_nominal_lin}
\end{multline}
\normalsize
where $\mathbf Z_G^{(r)}$ and $\mathbf Z_{B,q}^{(r)}$ are the values of $\mathbf Z_G$ and $\mathbf Z_{B,q}$ at the previous iteration $r$, respectively. Since both $\mathbf Z_G(\mathbf \Phi_l)$ and $\mathbf Z_{B,q}(\mathbf \Phi_l)$ are affine in $\mathbf \Phi_l$, the quantity $\psi_{q,j}^{(r)}(\mathbf \Phi_l)$ is affine in $\mathbf \Phi_l$. In addition, the values of the $\eta_{q,j}$ and $\beta_q$ are calculated based on the previous iteration matrices as $\eta_{q,j}^{(r)} = \left\| \mathbf \Psi_{q,j}^{(r)} \right\|_F$  and $\beta_q^{(r)} = \|\mathbf Z_{B,q}^{(r)}\|_2 \|\mathbf Z_G^{(r)}\|_2$. Then, the robust upper and lower bounds become

\small
\begin{align}
\overline{\psi}_{q,j}^{(r)}(\mathbf \Phi_l)
&=
|\alpha_j|^2
\left(
\psi_{q,j}^{(r)}(\mathbf \Phi_l)
+
2\varepsilon_{A_j}\eta_{q,j}^{(r)}
+
\varepsilon_{A_j}^2 \beta_q^{(r)}
\right),
\quad j\neq q,
\label{eq:upper_psi_sca}
\\
\underline{\psi}_{q,q}^{(r)}(\mathbf \Phi_l)
&=
|\alpha_q|^2
\left(
\psi_{q,q}^{(r)}(\mathbf \Phi_l)
-
2\varepsilon_{A_q}\eta_{q,q}^{(r)}
-
\varepsilon_{A_q}^2 \beta_q^{(r)}
\right).
\label{eq:lower_psi_sca}
\end{align}
\normalsize

Accordingly, the sensing constraint is approximated by
\begin{equation}
\sum_{j\neq q} \overline{\psi}_{q,j}^{(r)}(\mathbf \Phi_l)
-
\bar{\varrho}\, \underline{\psi}_{q,q}^{(r)}(\mathbf \Phi_l)
+
\sigma_B^2 \operatorname{Tr}(\mathbf U_q)
\le 0,
\qquad \forall q\in\mathcal Q.
\label{eq:sensing_surrogate_sca}
\end{equation}

After adding the penalty term $\Big[\zeta_3
\left(
\operatorname{Tr}(\mathbf \Phi_l)-(\pmb{\phi}_l^{(r)})^H \mathbf \Phi_l \pmb{\phi}_l^{(r)}
\right)\Big]$ to the objective function to satisfy the rank-one constraint, for a fixed trial value $\bar{\varrho}$, the $l$-th layer phase update is obtained by solving
\begin{subequations}
\begin{align}
\min_{\mathbf \Phi_l}
\quad
& \bar{\varrho} +
\zeta_3
\left(
\operatorname{Tr}(\mathbf \Phi_l)
-
(\boldsymbol{\phi}_l^{(r)})^H \mathbf \Phi_l \boldsymbol{\phi}_l^{(r)}
\right)
\\
\text{s.t.}\quad
&
\mathbf \Phi_l \succeq \mathbf 0,
\qquad
\operatorname{diag}(\mathbf \Phi_l)=\mathbf 1,
\\
&\eqref{Eq:robust_sinr_lmi},~\eqref{Eq:robust_covert_lmi},~\text{and}~
\eqref{eq:sensing_surrogate_sca}.
\end{align}
\label{eq:phase_tight_surrogate_problem}
\end{subequations}
\vspace{-0.5cm}

Problem \eqref{eq:phase_tight_surrogate_problem} is a convex semidefinite program. Compared with the LMI-based solution, this formulation offers a much lower complexity because it avoids constructing the large matrix $\mathbf X_q$ and its associated semidefinite constraint. Hence, the proposed norm-based robust surrogate provides an effective compromise between robustness and computational tractability in the layer-by-layer SIM phase optimization.

\subsection{Complete Algorithm, Convergence, and Computational Complexity}

Based on the previously formulated sub-problems and their corresponding solutions, the original problem~\eqref{Eq:Op3} is solved using an AO framework, where the convexified receive beamformers, transmit beamformers, and SIM phase shifts sub-problems are successively optimized while fixing the remaining variable blocks. 
Since SCA approximations and rank-one penalty-based reformulations are used, global optimality cannot be guaranteed. Nevertheless, following the standard convergence properties of AO/SCA methods, the proposed algorithm converges to a stationary solution. Since the auxiliary variable $\bar{\varrho}=1/\varrho$ is handled through bisection in all three sub-problems, each AO step involves a bisection search whose feasibility is determined by solving the corresponding convex SDP. The overall procedure is summarized in Algorithm~\ref{alg:AO}.

\begin{algorithm}
\small
\caption{Proposed AO Algorithm for Robust SIM-Assisted ISACC Framework}
\label{alg:AO}
\begin{algorithmic}[1]

\STATE \textbf{Input:} 
$\{\hat{\mathbf h}_k\}_{k\in\mathcal K}$,
$\{\hat{\mathbf a}_q\}_{q\in\mathcal Q}$, $\{\hat{\mathbf A}_q\}_{q\in\mathcal Q}$
$\{\mathbf{\Omega}^l\}_{l\in\mathcal L}$,
$\mathbf{\Omega}_t^1$, $\mathbf{\Omega}_r^1$,
$P_{\max}$, $\Gamma_k$, $\varpi$,
$\{\varepsilon_{h_k}\}$, $\{\varepsilon_{a_q}\}$,
$\{\varepsilon_{A_q}\}$, $\sigma_B^2$,
$\zeta_1$, $\zeta_2$, and $\zeta_3$.

\STATE \textbf{Initialize:}
$\{\mathbf W_k^{(0)}\}_{k\in\mathcal K}$,
$\mathbf R_0^{(0)}$,
$\{\mathbf U_q^{(0)}\}_{q\in\mathcal Q}$,
$\{\boldsymbol{\Theta}^{l,(0)}\}_{l\in\mathcal L}$,
and set $r\leftarrow1$.

\REPEAT

    \STATE \textbf{Receive beamforming:}
    Perform bisection over $\bar{\varrho}$ and solve~\eqref{eq:rx_sca_obj} for each trial value to obtain
    $\{\mathbf U_q^{(r)}\}_{q\in\mathcal Q}$, and recover $\{\mathbf u_q^{(r)}\}$.

    \STATE \textbf{Transmit beamforming:}
    Perform bisection over $\bar{\varrho}$ and solve~\eqref{Eq:Op5} for each trial value to obtain
    $\{\mathbf W_k^{(r)}\}_{k\in\mathcal K}$, $\mathbf R_0^{(r)}$, and recover $\{\mathbf w_k^{(r)}\}$.

    \STATE \textbf{SIM phase shift optimization:}
    Perform bisection over $\bar{\varrho}$.

    \REPEAT
        \FOR{$l=1,\ldots,L$}
            \STATE Solve~\eqref{Eq:phase_shift_prob} for the LMI-based design, or~\eqref{eq:phase_tight_surrogate_problem} for the low-complexity design, to obtain $\mathbf{\Phi}_l^{(r)}$.
            \STATE Recover $\boldsymbol{\theta}_l^{(r)}$ and set
            $\boldsymbol{\Theta}^{l,(r)}
            =\operatorname{diag}(\boldsymbol{\theta}_l^{(r)})$.
        \ENDFOR
    \UNTIL{the layer-sweep convergence criterion is satisfied.}

    \STATE Update all optimization and auxiliary variables.
    \STATE $r\leftarrow r+1$.

\UNTIL{the AO convergence criterion is satisfied.}

\STATE \textbf{Output:}
$\{\mathbf w_k^\ast\}_{k\in\mathcal K}$,
$\mathbf R_0^\ast$,
$\{\mathbf u_q^\ast\}_{q\in\mathcal Q}$,
$\{\boldsymbol{\Theta}^{l,\ast}\}_{l\in\mathcal L}$,
and $\varrho^\ast$.

\end{algorithmic}
\end{algorithm}

The computational complexity of Algorithm~\ref{alg:AO} is mainly determined by the SDPs solved within the receive beamforming, transmit beamforming, and SIM phase shift optimization steps. For the receive beamforming sub-problem~\eqref{eq:rx_sca_obj}, there are $Q$ positive semidefinite matrices of dimension $N_r$ and $Q$ reduced robust sensing LMIs of dimension $Qr_GN_r+1$. Hence, 
the corresponding computational complexity is $\mathcal{O}\left(
I_{\mathrm{B,R}}\left(D_{R}\right)^{3.5}
\right)$, where $I_{\mathrm{B,R}}$ is the number of bisection iterations and $D_{R} = QN_r+Q(Qr_GN_r+1)$. Similarly, the computational complexity of the transmit beamforming sub-problem~\eqref{Eq:Op5}
can be given as $\mathcal{O}\bigg(
I_{\mathrm{B,T}}\big(D_{T}\big)^{3.5}
\bigg)$, where $I_{\mathrm{B,T}}$ is the corresponding number of bisection iterations and $D_{T} = (K+1)N_t
+
(K+Q)(M+1)
+
Q(Qr_G+1)$. For the LMI-based SIM phase shift sub-problem~\eqref{Eq:phase_shift_prob}, 
the complexity is $\mathcal{O}\bigg(
I_{\mathrm{B,P}}I_{\mathrm S}L
\big(D_P^{\text{LMI}}\big)^{3.5}
\bigg)$, where $I_{\mathrm{B,P}}$ denotes the number of bisection iterations, $D_P^{\text{LMI}} = M
+
(K+Q)(M+1)
+
Q(QM^2+1)$, $L$ is the number of SIM layers that are successively optimized over $I_{\mathrm S}$ layer sweeps for every bisection trial. For the low-complexity phase shift formulation~\eqref{eq:phase_tight_surrogate_problem}, the $Q$ high-dimensional sensing LMIs are replaced by $Q$ scalar robust surrogate constraints. Therefore, the computational complexity is characterized as $\mathcal{O}\bigg(
I_{\mathrm{B,P}}I_{\mathrm S}L
\big(D_P^{\text{LC}}\big)^{3.5}
\bigg)$, where $D_P^{\text{LC}} = M
+
(K+Q)(M+1)$.

Consequently, if $I_{\mathrm{AO}}$ denotes the number of outer AO iterations, the overall computational complexity when adopting the LMI-based phase shift design is

\small
\begin{dmath}
\mathcal{O}\left(
I_{\mathrm{AO}}
\left[
I_{\mathrm{B,R}} \left(D_{R}\right)^{3.5}
+
I_{\mathrm{B,T}} \left(D_{T}\right)^{3.5}
+
I_{\mathrm{B,P}}I_{\mathrm S}L
\left(D_P^{\text{LMI}}\right)^{3.5}
\right]
\right),
\end{dmath}
\normalsize
whereas the proposed low-complexity phase shift design reduces it to

\small
\begin{dmath}
\mathcal{O}\left(
I_{\mathrm{AO}}
\left[
I_{\mathrm{B,R}} \left(D_{R}\right)^{3.5}
+
I_{\mathrm{B,T}}\left(D_{T}\right)^{3.5}
+
I_{\mathrm{B,P}}I_{\mathrm S}L
\left(D_P^{\text{LC}}\right)^{3.5}
\right]
\right),
\end{dmath}
\normalsize

The proposed dimensionality-reduction procedures therefore substantially decrease the computational burden of all three optimization blocks. In particular, the robust sensing LMI of dimension $QM^2+1$ is exactly reduced to $Qr_GN_r+1$ and $Qr_G+1$ in the receive and transmit beamforming sub-problems, respectively. Moreover, the low-complexity phase shift formulation completely avoids the $QM^2+1$-dimensional sensing LMI.

\section{Simulation Results}
\label{results}

\begin{table*}[t]
	\renewcommand{\arraystretch}{1.1}
        \centering
        \caption{Simulation Parameters}
	\begin{tabularx}{\textwidth}{Xc||Xc}
		\hline\hline
		\textbf{Parameter} & \textbf{Value} & \textbf{Parameter} & \textbf{Value} \\
		\hline
        Number of Tx antennas, $N_t$ & $6$ & Number of covert UEs, $K$  & $3$\\
		Number of Rx antennas, $N_r$ & $6$ & Number of Targets/wardens, $Q$ & $2$\\
		Number of SIM layers, $L$ & $3$ & Number of meta-atoms per layer, $M$ &  $25$ \\
		Carrier frequency & $10$ GHz & Communication bandwidth, $B_w$ & $1$ MHz \\
		BS height, $Z$  & $15$ m& Radius of service area & $20$ m\\
		Maximum allowed transmit power $P_{\text{max}}$ & $30$-$40$ dBm & Covertness threshold $\epsilon$  & $0.1$\\
		Users' rate requirement, $R_k^{\text{min}}$ & $0.1$ bps/Hz &  Number of Warden observations per block, $T$ & $1000$\\
        meta-atom spacing, $d_{\text{element}}$ &$\lambda/2$ & Area of each meta-atom, $A_e$ & $\lambda^2/4$\\
        Noise power spectral density & $-174$ dBm/Hz& SIM layers spacing, $d_{\text{SIM}}$& $\lambda/4$\\
		\hline
	\end{tabularx}
	\label{tab1}
\end{table*}

In this section, comprehensive numerical simulations are conducted to evaluate the performance of the proposed robust SIM-assisted ISACC framework. The system operates in a three-dimensional Cartesian coordinate system with the BS located at $\left(0,0,Z\right)$, where $Z$ represents the BS height. Both downlink communication UEs and targets are randomly generated following a uniform distribution within a $20$ m radius circle centered at $\left(20,20,0\right)$ m. The inter-layer spacing is set to $d_{\text{SIM}} = \frac{\lambda}{4}$~\cite{wang2024multi}, while the spacing between two adjacent meta-atoms is set to $d_{\text{element}} = \frac{\lambda}{2}$. We assume a path loss exponent of $\eta=2.5$ and a Rician factor of $\kappa = 10$ dB. The penalty factors are set to $\zeta_1=\zeta_2=\zeta_3=10^{-2}$, while the CSI bounds are set to $10^{-3}$. Unless otherwise stated, the main simulation parameters are given in Table~\ref{tab1}. The propagation distance, $r_{m,\tilde{m}}^l$, from the $\tilde{m}$-th meta-atom in the $l-1$-th layer to the $m$-th meta-atom in the $l$-th layer is given by

\small
\begin{dmath}
    r_{m,\tilde{m}}^l = \sqrt{d_{\text{SIM}}^2+ d_{\text{element}}^2 \left[\left( i_x\left(m\right)-i_x\left(\tilde{m}\right)\right)^2+\left( i_z\left(m\right)-i_z\left(\tilde{m}\right)\right)^2\right]},
\end{dmath}
\normalsize
where $i_x(m) = \text{mod}(m-1, M_x) + 1$ and $i_z(m) = \left\lceil \frac{m}{M_x} \right\rceil$ represent the 2D grid column and row indices of the $m$-th meta-atom. Additionally, the propagation distance, $r_{m,n_t}^1$, from the $n_t$-th transmit antenna to the $m$-th meta-atom in the $1^{st}$ layer is given by~\eqref{Eq:45} at the top of this page. Similar to~\eqref{Eq:45}, we can get the propagation distance, $r_{m,n_r}^1$, from the $m$-th meta-atom in the $1^{st}$ layer to the $n_r$-th receive antenna.
\begin{figure*}[!t]
\begin{equation}
     r_{m,n_t}^1 = \sqrt{d_{\text{SIM}}^2+ d_{\text{element}}^2 \left[ \left(\text{mod}\left( m-1,M_x\right)-\frac{M_x-1}{2}\right)-\left (n_t- \frac{1+N_t}{2} \right )\right]^2 +  d_{\text{element}}^2\left(\left \lceil \frac{m}{M_x} \right \rceil - \frac{1+M_z}{2} \right)^2} \label{Eq:45}
\end{equation}
\vspace*{2pt}
\hrulefill
\end{figure*}

To show the effectiveness of our proposed robust SIM-assisted ISACC framework, we compare the achieved performance with four different baseline schemes. More specifically, we consider: \textbf{1) Sensing-Centric Scheme}: which removes the communication QoS constraints to evaluate the maximum achievable target sensing SINR without supporting communication users; \textbf{2) No-Covertness Scheme}: which relaxes the covertness detection constraints to quantify the performance trade-offs and penalties incurred by enforcing strict covert transmission; \textbf{3) Fixed (Random) Phase Shift Scheme}: where the wave-domain phase shifts are randomly initialized and held constant, isolating the performance gain directly contributed by joint digital and wave-domain optimization; and \textbf{4) Fully Digital Beamforming Scheme (No SIM)}: which removes the SIM architecture entirely and relies exclusively on conventional digital-domain processing to meet the target objectives and constraints. 

\begin{figure}[!t]
\centerline{\includegraphics[width=0.35\textwidth]{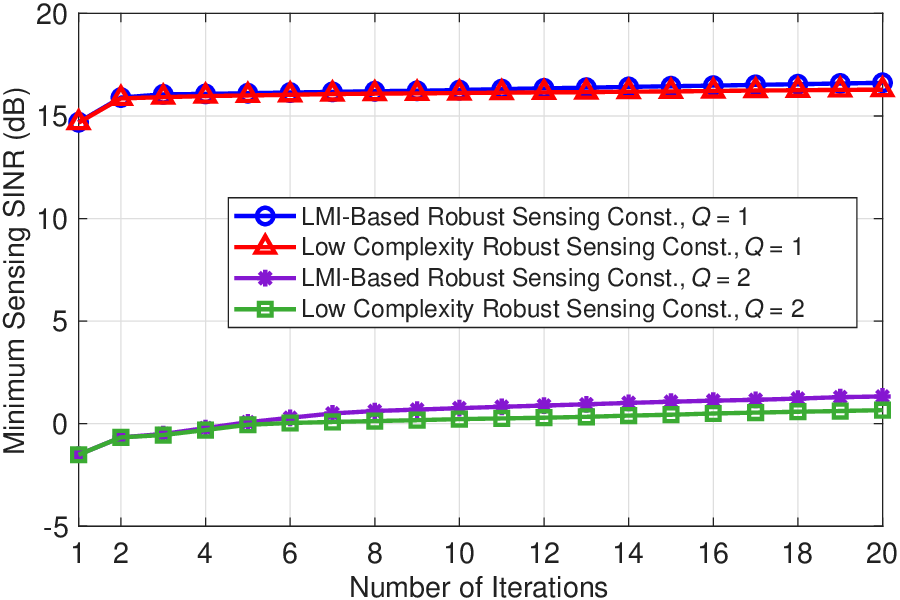}}
\caption{Convergence behavior comparison of the proposed LMI-based and low-complexity phase shift designs in terms of minimum sensing SINR at $P_{\text{max}} = 35$ dBm and $M = 3\times3$ meta-atoms per layer for $Q\in\left\{1,2\right\}$ sensing targets.}
\label{res1}
\end{figure}

The convergence performance of the proposed robust SIM-assisted ISACC system is presented in Fig.~\ref{res1}. It compares the LMI-based and low-complexity phase shift designs for different number of targets. Due to the high computational cost of the LMI-based design, we consider the special case of a SIM with only 9 meta-atoms in each layer arranged in a $3\times3$ array. As shown, both approaches achieve almost the same performance, with the low-complexity design substantially reduces the computational complexity. For $Q=1$, the SINR approaches a stable value within a few iterations, whereas the improvement is more gradual for $Q=2$. Increasing the number of targets from one to two substantially reduces the minimum sensing SINR. This drop reflects the limited spatial DoF provided by the small SIM array, which restricts its ability to support multiple sensing targets. Given the comparable performance of both approaches, we will present only the low-complexity design in all subsequent simulations.

\begin{figure}[!t]
\centerline{\includegraphics[width=0.35\textwidth]{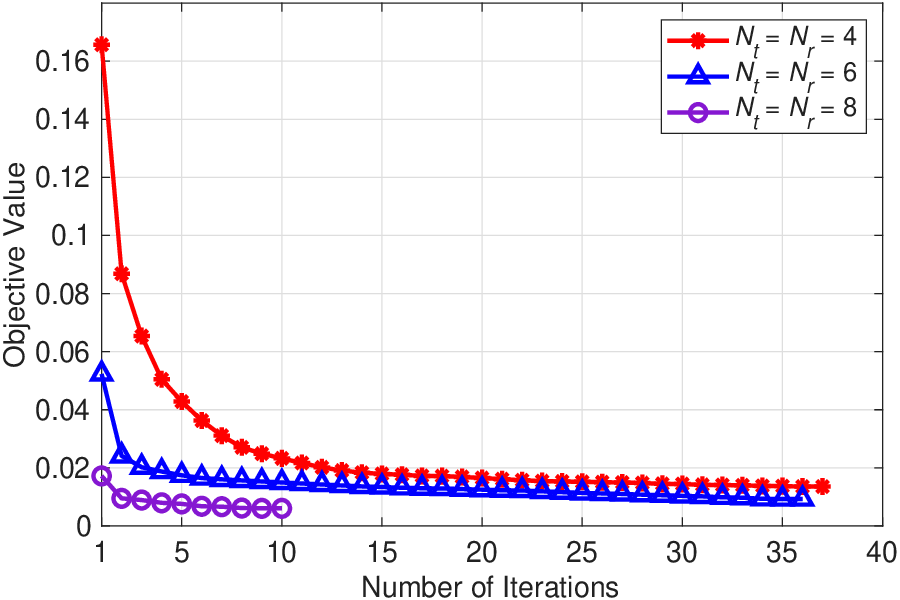}}
\caption{Convergence behavior of the optimization problem objective $\bar{\varrho}$ for different antenna configurations, $N_t=N_r\in\{4,6,8\}$.}
\label{res2}
\end{figure}

\begin{figure}[!t]
\centerline{\includegraphics[width=0.35\textwidth]{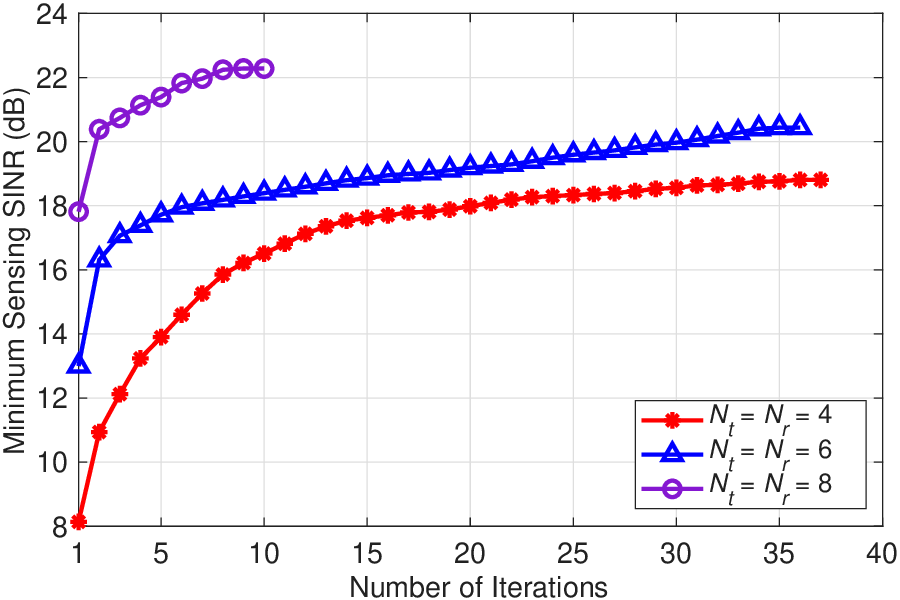}}
\caption{Convergence behavior of the actual minimum sensing SINR objective in dB for different antenna configurations, $N_t=N_r\in\{4,6,8\}$.}
\label{res3}
\end{figure}

Figures~\ref{res2} and~\ref{res3} show the convergence behavior of the proposed SIM-assisted ISACC system for different numbers of transmit/receive antennas. Fig.~\ref{res2} presents the objective value $\bar{\varrho}$, while Fig.~\ref{res3} shows the corresponding actual minimum sensing SINR in dB. For $N_t=N_r=8$, both curves become nearly stable after approximately 8–10 iterations. The cases with $N_t=N_r=4$ and $N_t=N_r=6$ improve more gradually until they converge at approximately 35 iterations. Increasing the number of antennas also leads to a lower final objective value and a higher minimum sensing SINR. Specifically, the achieved minimum sensing SINR values are approximately $18.8$, $20.6$, and $22.3$ dB for $N_t=N_r=4$, $6$, and $8$, respectively. This improvement is consistent with the additional spatial DoF available for transmit and receive beamforming. More antennas provide greater flexibility to strengthen the desired sensing signals and suppress interference, thereby improving the sensing performance of the weakest target.

\begin{figure}[!t]
\centerline{\includegraphics[width=0.35\textwidth]{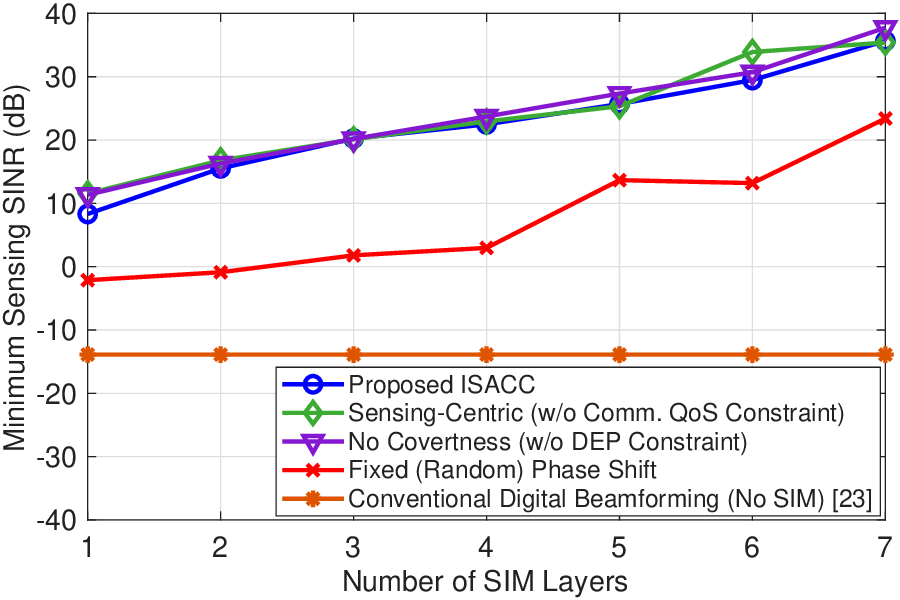}}
\caption{Effect of increasing the number of SIM layers on the achieved sensing performance for the proposed scheme as well as other benchmarks.}
\label{res4}
\end{figure}

The effect of the number of SIM layers on the minimum sensing SINR is presented in Fig.~\ref{res4} for the proposed scheme as well as other benchmarks. In general, the sensing performance improves as more layers are added. This improvement results from the additional spatial DoF provided by the SIM, which enables more effective wave-domain beamforming and interference control. The achieved performance by the proposed ISACC framework increases from approximately $8$ dB with one layer to $35$ dB with seven layers. Even with a single SIM layer, it outperforms conventional digital beamforming without a SIM, which provides a constant sensing SINR of about $-14$ dB. The gain is approximately $22$ dB at $L=1$ and increases to nearly $50$ dB at $L=7$. Moreover, the proposed design also consistently outperforms the fixed random phase shift scheme. For example, at $L=1$, the proposed method achieves about $8$ dB, compared with approximately $-2$ dB for random phase shifts, corresponding to a gain of nearly $10$ dB.  At $L=7$, the proposed design gives a performance of $12$ dB above the random-phase benchmark. These results show that optimizing the SIM phases is important. Nevertheless, the random-phase scheme also benefits from increasing the number of layers, since additional layers provide extra spatial DoF even when their phase shifts are not optimized.

Additionally, the sensing-centric and no-covertness schemes represent upper bounds on the achieved sensing performance since they both relax one of the constraints form the formulated problem, where the former removes the communication QoS constraints, while the latter removes the detection error probability constraint. As expected, both relaxed schemes generally achieve higher sensing SINR than the fully constrained ISACC design, although their values are close to the proposed result for some layer numbers.

\begin{figure}[!t]
\centerline{\includegraphics[width=0.35\textwidth]{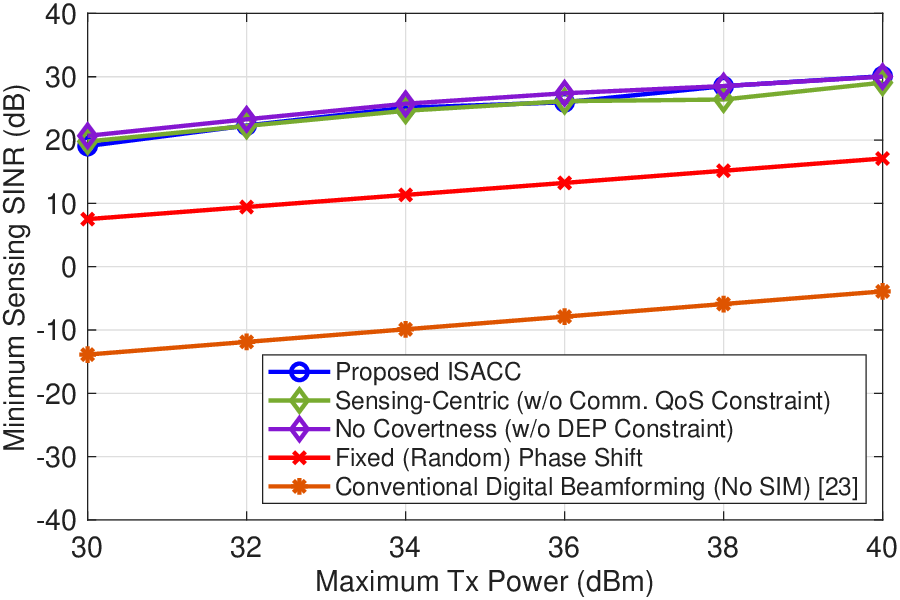}}
\caption{Effect of increasing the maximum transmit power on the achieved sensing performance for the proposed scheme as well as other benchmarks.}
\label{res5}
\end{figure}

Figure~\ref{res5} presents the achieved minimum sensing SINR against the maximum transmit power for the proposed SIM-assisted ISACC system as well as other benchmarks. Across all schemes, the sensing SINR increases monotonically with power, as higher transmit budgets directly yield stronger target echo signals. As mentioned previously, since the sensing-centric and no-covertness baselines relax the communication QoS and covertness constraints, respectively, they represent theoretical upper bounds on the system performance. Again, it is apparent that the proposed scheme operates very close to these unconstrained bounds across the entire power range despite enforcing all constraints. Furthermore, it substantially outperforms both the fixed random phase shift baseline and conventional digital beamforming without a SIM. Specifically, at $P_{\text{max}} = 40$ dBm, the proposed design achieves a sensing SINR approximately $13$ dB higher than the random phase shift approach and $34$ dB higher than conventional digital beamforming. These gains highlight the structural advantage of the SIM architecture and the critical importance of optimizing its phase shifts.

\begin{figure}[!t]
\centerline{\includegraphics[width=0.35\textwidth]{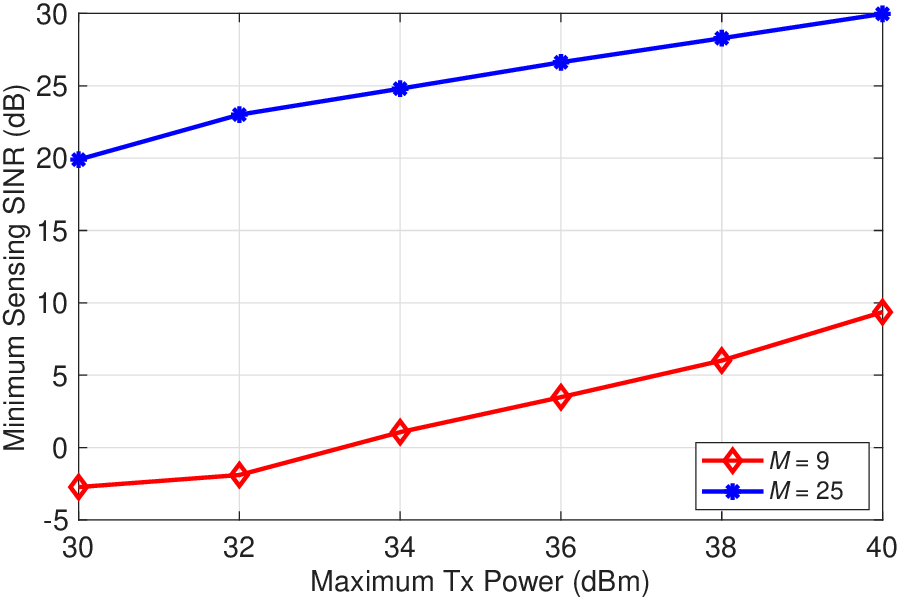}}
\caption{Impact of the number of meta-atoms per layer, $M$, on the minimum sensing SINR under varying transmit power budgets.}
\label{res6}
\end{figure}

Figure~\ref{res6} illustrates how the number of meta-atoms per layer, $M$, influences the minimum sensing SINR as a function of the maximum transmit power budget. It is clear that scaling up $M$ from $9$ to $25$ substantially raises the sensing SINR across all transmit power levels. This behavior is attributed to the fact that a larger $M$ provides finer spatial resolution and greater wave-manipulation capability, enabling tighter beam focusing toward target directions while maintaining compliance with communication and covertness constraints. In particular, at $P_{\text{max}} = 40$ dBm, the setup with $M = 25$ achieves an SINR of approximately $30$ dB, compared to just $9.5$ dB for $M = 9$. This degradation highlights the limited spatial DoF offered by a smaller SIM array, which restricts its ability to form directional beams and support multiple communication users and multiple sensing targets simultaneously. Moreover, this large performance gap shows that expanding the SIM size offers a powerful means of compensating for low transmit power budgets.

\begin{figure}[!t]
\centerline{\includegraphics[width=0.35\textwidth]{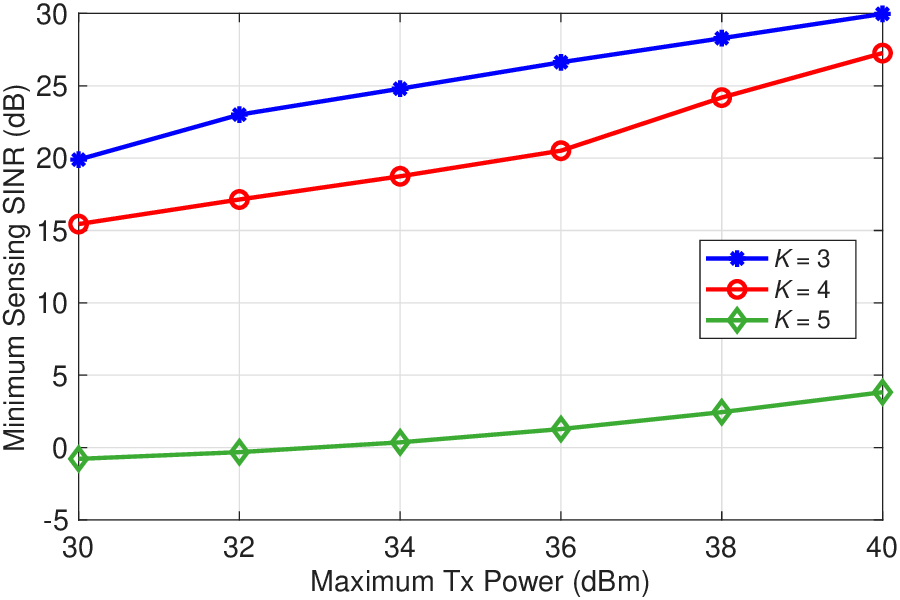}}
\caption{Impact of the number of communication UEs, $K$, on the minimum sensing SINR under varying transmit power budgets.}
\label{res7}
\end{figure}

The minimum sensing SINR versus the maximum transmit power $P_{\text{max}}$ is presented in Fig.~\ref{res7} for different numbers of communication users $K\in\{3,4,5\}$. As expected, the sensing SINR increases monotonically with power for all configurations. However, increasing the number of communication users $K$ leads to a notable degradation in sensing performance. This trade-off occurs because serving more communication users requires allocating a larger portion of the power budget and spatial DoF to satisfy their respective QoS constraints, thereby leaving limited resources for target sensing. For example, at $P_{\text{max}} = 40$ dBm, the minimum sensing SINR drops from approximately $30$ dB for $K = 3$ to $27.2$ dB for $K = 4$, and falls sharply to $3.8$ dB when $K = 5$. These results clearly demonstrate the fundamental trade-off between communication and sensing in multi-user ISAC systems.

\section{Conclusion}
\label{sec_conc}

This paper developed a robust transceiver design for a multi-user, multi-target SIM-assisted ISACC system under bounded CSI uncertainty. The proposed AO framework jointly optimizes the transmit and receive beamformers, sensing covariance matrix, and SIM phases to maximize the worst-case minimum sensing SINR while satisfying the communication QoS, power budget, SIM phase shifts, and covertness constraints. Exact column-space reductions were derived for the robust sensing LMIs in the transmit- and receive-beamformer updates, while a norm-based phase-update method was introduced to reduce the computational cost. Numerical results show that at the low-complexity phase update closely matches the LMI-based design at substantially lower complexity. Furthermore, at $P_{\max}=40$ dBm, the proposed ISACC framework outperforms the random-phase and the fully-digital beamforming benchmarks by approximately $13$ and $34$ dB, respectively. In addition, increasing the number of SIM layers from $1$ to $7$ raises the sensing SINR from approximately $8$ to $35$ dB, which reflects the potential of SIMs for robust covert ISAC. However, increasing the number of users from $3$ to $5$ reduces the sensing SINR from approximately $30$ to $4$ dB at $P_{\max}=40$ dBm, which shows the tradeoff between sensing and communications.
{\appendix[Proof of Proposition~\ref{thm:dimension_reduction}]
\label{App1}

To significantly reduce the computational complexity while preserving the exact structure of the robust LMI, we exploit the fact that $\mathbf G \in\mathbb{C}^{M\times N_t}$ has a low-dimensional column space since generally $N_t < M$. Let $r_G=\operatorname{Rank}(\mathbf G)$ and consider the compact singular value decomposition (SVD) of $\mathbf G = \mathbf U_G\mathbf \Sigma_G\mathbf V_G^H$, where $\mathbf U_G\in\mathbb C^{M\times r_G}$ contains an orthonormal basis spanning the column space of $\mathbf G$. Let $\mathbf G\mathbf R_x\mathbf G^H = \mathbf U_G\mathbf T\mathbf U_G^H$, where the reduced-dimensional matrix $\mathbf T = \mathbf U_G^H \mathbf G\mathbf R_x\mathbf G^H \mathbf U_G \in\mathbb C^{r_G\times r_G}$. Consequently, $\left( \mathbf G\mathbf R_x\mathbf G^H \right)^T = \mathbf U_G^* \mathbf T^T \mathbf U_G^T$.

Using the mixed-product property of the Kronecker product, each $\mathbf Y_{q,j}$ can therefore be factorized as
\begin{equation}
\mathbf Y_{q,j}
=
\mathbf C_j
\left(
\mathbf T^T\otimes\mathbf U_q
\right)
\mathbf C_j^H,
\end{equation}
where $\mathbf C_j
=
|\alpha_j|
\left(
\mathbf U_G^*
\otimes
\mathbf B
\right)$. Since for this sub-problem $\mathbf R_x$ is fixed, while $\mathbf U_q$ is the optimization variable, $\mathbf C_j$ is constant with respect to the receive beamforming variables. Hence, defining $\mathbf C = \operatorname{diag}
\left(
\mathbf C_1,\ldots,\mathbf C_Q
\right)$, the original high-dimensional matrix $\mathbf X_q$ can be represented exactly as
\begin{equation}
\mathbf X_q
=
\mathbf C
\mathbf K_q
\mathbf C^H,
\end{equation}
where $\mathbf K_q
=
\operatorname{diag}
\left(
\mathbf K_{q,1},\ldots,\mathbf K_{q,Q}
\right)$, with $\mathbf K_{q,j}
=
\mathbf T^T\otimes\mathbf U_q,~ j\neq q$ and $\mathbf K_{q,q}
=
-\bar{\varrho}
\left(
\mathbf T^T\otimes\mathbf U_q
\right)$.

For further reduction, we perform the compact QR factorization $\mathbf C
=
\mathbf U_C\mathbf R_C$, where $\mathbf U_C^H\mathbf U_C=\mathbf I$. Accordingly, $\mathbf X_q$ can be rewritten as
\begin{equation}
\mathbf X_q
=
\mathbf U_C
\widetilde{\mathbf K}_q
\mathbf U_C^H,
\end{equation}
where $\widetilde{\mathbf K}_q
=
\mathbf R_C
\mathbf K_q
\mathbf R_C^H$. Furthermore, the reduced-dimensional nominal uncertainty vector is now expressed as $\widetilde{\mathbf g}_A
=
\mathbf U_C^H
\hat{\mathbf g}_A$. Consequently, the original robust LMI can be represented in the reduced-dimensional space as in~\eqref{Eq:red_LMI}.
This completes the proof. $\blacksquare$
}


\bibliographystyle{IEEEtran}
\bibliography{References}

 




\end{document}